\documentclass[a4paper,fleqn,usenatbib]{mnras}
\usepackage[T1]{fontenc}
\usepackage{ae,aecompl}
\usepackage{graphicx}	
\usepackage{amsmath}	
\usepackage{amssymb}	
\usepackage{amsbsy}
\usepackage{float}
\usepackage{color}
\usepackage{txfonts}

\usepackage{mathrsfs,bm}
\newcommand{\bcdot}{\ensuremath{%
  \mathchoice%
   {\mskip\thinmuskip\lower0.2ex\hbox{\scalebox{1.5}{$\cdot$}}\mskip\thinmuskip}}%
   {\mskip\thinmuskip\lower0.2ex\hbox{\scalebox{1.5}{$\cdot$}}\mskip\thinmuskip}%
   {\lower0.3ex\hbox{\scalebox{1.2}{$\cdot$}}}%
   {\lower0.3ex\hbox{\scalebox{1.2}{$\cdot$}}}%
}

\hypersetup{draft}

\title[Magnetic fields in disk galaxies]{Magnetic field formation in the Milky Way-like disk galaxies of the Auriga project}

\author[R.~Pakmor et al.]  {R\"udiger~Pakmor$^1$\thanks{E-mail: ruediger.pakmor@h-its.org}, Facundo A. G\'omez$^2$, Robert J.~J. Grand$^{1,3}$, Federico Marinacci$^4$, \newauthor Christine M. Simpson$^1$, Volker Springel$^{1,3}$, David J.~R. Campbell$^5$, Carlos S. Frenk$^5$, \newauthor Thomas Guillet$^{1,3}$, Christoph Pfrommer$^1$, Simon D.~M. White$^2$
  \vspace*{0.2cm}  \\
  $^1$Heidelberger Institut f\"{u}r Theoretische Studien,
  Schloss-Wolfsbrunnenweg 35, 69118 Heidelberg, Germany\\
  $^2$Max-Planck-Institut f\"ur Astrophysik, Karl-Schwarzschild-Str. 1, D-85748, 
Garching, Germany\\
  $^3$Zentrum f\"ur Astronomie der Universit\"at Heidelberg, ARI, M\"onchhofstrasse
12-14, 69120 Heidelberg, Germany\\
  $^4$Kavli Institute for Astrophysics and Space Research, 
  Massachusetts Institute of Technology, Cambridge, MA 02139, USA\\
  $^5$Institute for Computational Cosmology, Department of Physics, Durham University, 
  South Road, Durham, DH1 3LE, UK
}
  
\pubyear{2017}

\begin{document}

\label{firstpage}
\pagerange{\pageref{firstpage}--\pageref{lastpage}}

\maketitle

\begin{abstract}
The magnetic fields observed in the Milky~Way and nearby galaxies appear to be in equipartition with the turbulent, thermal, and cosmic ray energy densities, and hence are expected to be dynamically important. However, the origin of these strong magnetic fields is still unclear, and most previous attempts to simulate galaxy formation from cosmological initial conditions have ignored them altogether.  Here, we analyse the magnetic fields predicted by the simulations of the Auriga Project, a set of 30 high-resolution cosmological zoom simulations of Milky~Way-like galaxies, carried out with a moving-mesh magneto-hydrodynamics code and a detailed galaxy formation physics model.  We find that the magnetic fields grow exponentially at early times owing to a small-scale dynamo with an e-folding time of roughly $100\,\rm{Myr}$ in the center of halos until saturation occurs around $z=2-3$, when the magnetic energy density reaches about $10\%$ of the turbulent energy density with a typical strength of $10-50\,\rm{\mu G}$. In the galactic centers the ratio between magnetic and turbulent energy remains nearly constant until $z=0$. At larger radii, differential rotation in the disks leads to linear amplification that typically saturates around $z=0.5$ to $z=0$. The final radial and vertical variations of the magnetic field strength can be well described by two joint exponential profiles, and are in good agreement with observational constraints. Overall, the magnetic fields have only little effect on the global evolution of the galaxies as it takes too long to reach equipartition. We also demonstrate that our results are well converged with numerical resolution.

\end{abstract}


\begin{figure*}
  \centering
  \includegraphics[width=0.99\textwidth]{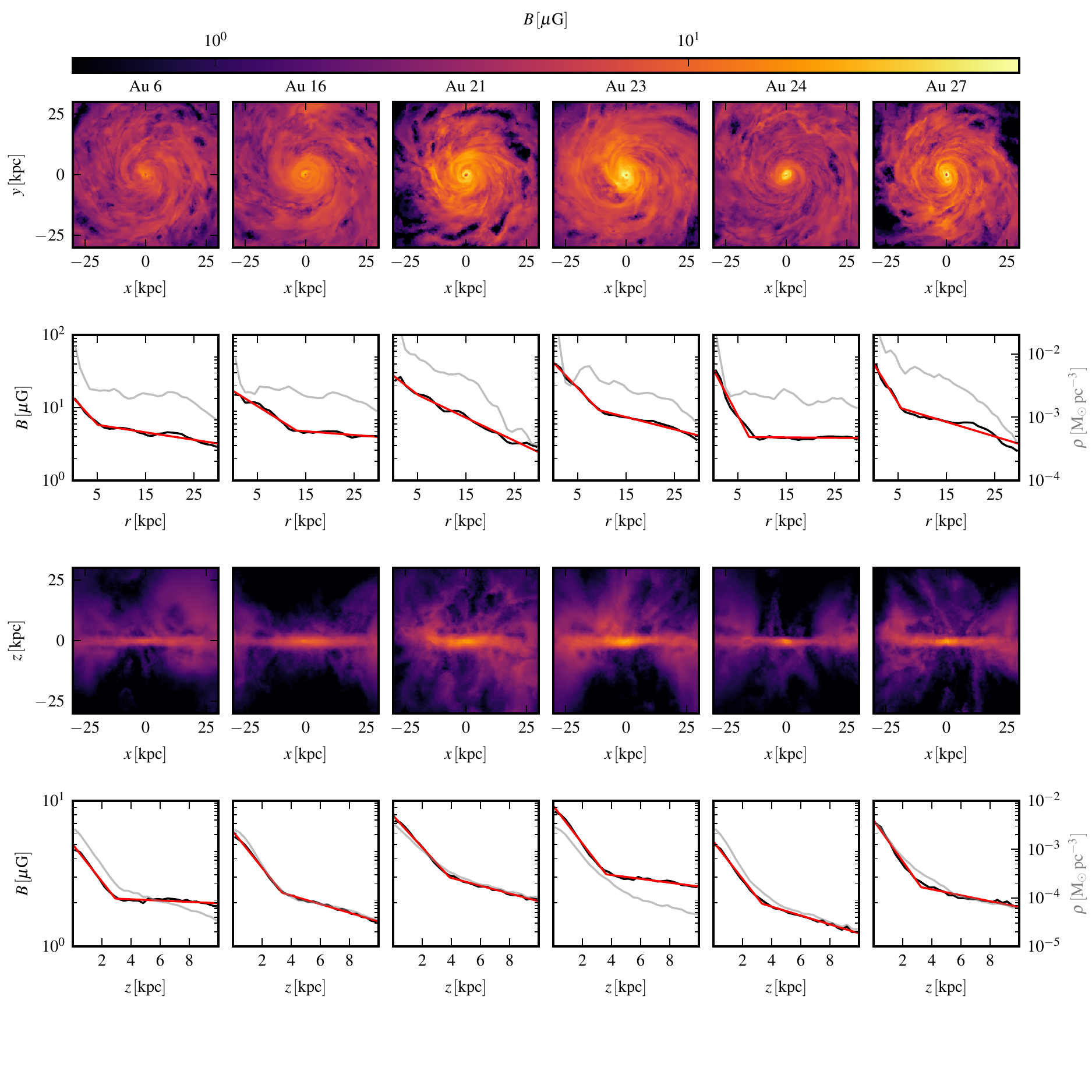}
  \caption{Magnetic field strength in the disks of 6 selected halos at $z=0$. The first row shows face-on projections with a depth of $2\,\rm{kpc}$ centered on the plane of the disk. The second row shows the radial profile of the magnetic field strength (black) and gas density (grey), measured in rings with a height of $2\,\rm{kpc}$ The red line is a double exponential fit to the magnetic field profile. The third row shows edge-on projections with a depth of $30\,\rm{kpc}$. The last row shows the vertical profile of the magnetic field strength in cylinders with a radius of $30\,\rm{kpc}$. The black, grey, and red lines display the profiles of magnetic field strength and gas density, respectively, and a double exponential fit to the magnetic field profile.}
  \label{fig:disks}
\end{figure*}

\section{Introduction}

Magnetic fields have been observed in the interstellar medium (ISM) for a large number of galaxies. One of the most important results is that the magnetic and turbulent energy densities in nearby galaxies are roughly in equipartition \citep{Boulares1990,Beck1996,Fletcher2010}. Therefore, at least today, magnetic fields are expected to be dynamically important for the evolution of galaxies. Moreover, strong magnetic fields have also been measured in galaxies at much earlier times, up to $z=2$ \citep{Bernet2008}. For an overview of different observational probes and their results see \citet{Beck2016}. In addition to their direct influence on gas dynamics, magnetic fields also play an important role in controlling the direction of anisotropic transport processes of charged particles, which is of immediate relevance for thermal conduction and the motion of cosmic rays.

Despite their importance, the origin and evolution of galactic magnetic fields is still not fully understood \citep{Kulsrud2008,Beck2016}. To reach their observed strength magnetic fields have to be amplified by several orders of magnitude from cosmological seed fields \citep{Durrer2013} or weak magnetic fields generated in stars and supernovae \citep{Bisnovatyi-Kogan1973,Hanayama2005}. The amplification process is highly non-linear and is inherently coupled to the complex baryonic physics that regulates star formation and the formation and evolution of galaxies. It is thus best studied by means of detailed numerical simulations that attempt to take all relevant processes into account. 

There is a long history of numerical simulations that study the amplification of magnetic fields in cosmological objects. Several full cosmological simulations of galaxy clusters \citep{Dolag1999, Dubois2008, Donnert2009,Vazza2014} and galactic halos \citep{Beck2012,Marinacci2015} have found that a turbulent dynamo can efficiently amplify magnetic fields in halos, although the amount of the reported amplification varies with numerical schemes and effective resolution. However, until recently cosmological MHD simulations were unable to form anything resembling realistic galaxies owing to shortcomings in the modelling of feedback processes and severe resolution limits.

Alternatively, numerical simulations of isolated galaxies that include magnetic fields also showed that the magnetic field can be efficiently amplified by a turbulent dynamo once a high enough resolution is available and sufficiently strong turbulence is present \citep{Wang2009,Hanasz2009,Dubois2010,Pakmor2013,Rieder2016,Pakmor2016b}. These idealised simulations were able to include more physical processes and reach significantly higher spatial resolution than their cosmological counterparts, but the lack of a consistent cosmological embedding and the arbitrariness of the initial conditions limit their predictive power. In particular, continued growth and accretion onto galaxies and the interaction with other halos and satellites play a significant role in shaping galaxies.

Utilising new modelling capabilities, we recently carried out the first ever high-resolution cosmological zoom-simulation of the formation of a Milky Way-like galaxy that included magnetic fields, showing that the observed strong large-scale magnetic fields can be naturally obtained from the amplification of a small cosmological seed field \citep{Pakmor2014}. In this paper, we significantly extend this pilot study by analysing the amplification of magnetic fields and their strengths at $z=0$ in the $30$ galaxies of the Auriga project \citep{Auriga}, a suite of high-resolution cosmological zoom simulations including a state of the art physics model and fully coupled magnetic fields. Here we focus on the overall evolution of the magnetic field strengths, and how it correlates with other galaxy properties.  An analysis of the structure and evolution of the magnetic field is addressed in a companion paper (Pakmor et al., in prep.).

This paper is structured as follows. We discuss the setup and physics model of our simulations in Sec.~\ref{sec:methods}. We then characterise the magnetic field strength in the galaxies at $z=0$ in Sec.~\ref{sec:disk} and analyse the amplification mechanisms that produce the final fields in Sec.~\ref{sec:amplification}. We briefly discuss the dynamical impact of the magnetic field and resolution effects in Sec.~\ref{sec:resolution}. Finally, we summarise our results in Sec.~\ref{sec:summary}.

\section{Methods and Setup}

\label{sec:methods}

The Auriga galaxies are formed in high-resolution zoom simulations of Milky Way-sized halos selected from the dark matter only simulation of the Eagle box \citep{Schaye2005Eagle} with a boxsize of $100\,\rm{Mpc}$. The halos are chosen to have a virial mass ($M_{200,\rm{crit}}$) between $10^{12}\,\rm{M_\odot}$ and $2 \times 10^{12}\,\rm{M_\odot}$ at $z=0$ and to fulfill a mild isolation criterion \citep{Auriga}.

The simulations employ the moving-mesh code \textsc{arepo} \citep{Arepo} that models dark matter, stars, and black holes as collisionless particles, and gas on an unstructured Voronoi-mesh that is advected with the flow and changes its structure with time, yielding a quasi Lagrangian description of hydrodynamics. Small correction motions are added where needed to preserve a reasonably regular mesh geometry, and local refinement and derefinement operations are applied whenever the mass content per cell deviates by more than a factor two from a target mass resolution. The equations of hydrodynamics are solved on this mesh using a second order finite volume scheme \citep{Pakmor2016}. The high resolution region around the target halo extends out to $4\,R_{200}$ of the halo in the parent DM-only simulation at $z=0$, and has a high resolution dark matter particle mass of $3 \times 10^5\,\rm{M_\odot}$ and a typical mass resolution for high resolution gas cells and star particles of $5 \times 10^4\,\rm{M_\odot}$. We use a constant comoving gravitational softening for collisionless particles of $250 h^{-1}\,\rm{pc}$ until $z=1$ and keep it constant in physical units thereafter. The gravitational softening of gas cells is set to the maximum of $2.8$ times their radius and the softening of collisionless particles.

Our model for baryonic physics consists of a slightly updated version of \citet{Marinacci2014}. It includes atomic and metal line cooling and a spatially uniform UV background \citep{Vogelsberger2013} and employs a sub-grid model for the interstellar medium and star formation \citep{Springel2003}. Stellar evolution is treated self-consistently and includes metal enrichment from core collapse supernovae, thermonuclear supernovae, and AGB stars \citep{Vogelsberger2013}. Feedback from core collapse supernovae is taken into account by a non-local effective kinetic wind model that isotropically injects momentum from the star-forming ISM and launches a wind just outside the star-forming ISM. The kinetic wind model is implemented using wind particles that are formed in star forming regions and recouple as soon as their density drops below $10\%$ of the star formation threshold. This usually happens within the disk and launches a galactic wind from the disk. We also include supermassive black holes and model accretion onto them and their feedback as AGN \citep{BH_paper}. For a detailed description of the physics model see \citet{Auriga}.

Magnetic fields are included in the limit of ideal magnetohydrodynamics (MHD). We employ the implementation of cell-centered magnetic fields in \textsc{arepo} \citep{Pakmor2011} that uses the HLLD Riemann solver \citep{miyoshi2005a} to compute fluxes and the Powell scheme \citep{Powell1999} for divergence cleaning \citep{Pakmor2013}. We start with a uniform magnetic seed field with comoving strength $10^{-14}\,\rm{G}$ at $z=127$ (equivalent to a physical strength of $2 \times 10^{-4} \mu \rm{G}$) oriented along the $z$-coordinate of the simulation box. This seed field strength is many order of magnitudes larger than plausible values for a cosmological seed field from inflation or fields seeded by Biermann batteries \citep{Kulsrud2008}. However, as shown in previous work, the information about the initial configuration and strength of the magnetic field is quickly erased by the exponential dynamo in collapsed halos \citep{Pakmor2014,Marinacci2015} and the initial strength is still small enough to be dynamically irrelevant outside collapsed halos \citep{Marinacci2016}.

We assume that the magnetic energy of gas that forms a star particle representing a population of stars is locked up in the star particle and thus removed from the gas \citep{Pakmor2014}. Similarly, the magnetic energy of gas that forms a wind particle is removed from the gas as well.

\begin{table*}
  \centering 
  \begin{tabular}{r r r r r r r r r r r}
  \hline\hline
  Halo & $B_\mathrm{center}$ & $B_\mathrm{disk}$ & $\rho_\mathrm{center}$ & D/T & $r^B_{0}$ & $r^B_\mathrm{inner}$ & $r^B_\mathrm{outer}$ & $h^B_{0}$ & $h^B_\mathrm{inner}$ & $h^B_\mathrm{outer}$ \\ 
   & $\mathrm{[\mu G]}$ & $\mathrm{[\mu G]}$ & $\mathrm{[M_\odot\,pc^{-3}]}$ & & $\mathrm{[kpc]}$ & $\mathrm{[kpc]}$ & $\mathrm{[kpc]}$ & $\mathrm{[kpc]}$ & $\mathrm{[kpc]}$ & $\mathrm{kpc}$ \\ 
  \hline
  Au-1 & 35.0 &  6.0 & $4.3\times10^{-2}$ & 0.40 &  7.8 &   6.0 & 13.1 &  4.4 &  4.6 & 17.7\\ 
  Au-2 & 31.9 &  5.5 & $2.1\times10^{-2}$ & 0.73 & 10.7 &   5.9 & 42.8 &  3.6 &  3.0 & 12.7\\ 
  Au-3 & 22.5 &  6.3 & $2.2\times10^{-2}$ & 0.76 & 13.7 &  10.2 & 42.6 &  3.0 &  3.3 & 34.0\\ 
  Au-4 & 72.9 &  8.7 & $7.8\times10^{-2}$ & 0.65 &  6.8 &   4.1 & 12.8 &  4.0 &  4.6 & 10.8\\ 
  Au-5 & 48.8 &  8.1 & $3.7\times10^{-2}$ & 0.71 &  6.5 &   5.6 & 11.5 &  3.2 &  3.4 & 18.5\\ 
  Au-6 & 13.9 &  4.3 & $1.3\times10^{-2}$ & 0.67 &  5.1 &   5.9 & 42.0 &  2.9 &  3.4 & 96.0\\ 
  Au-7 & 22.2 &  7.0 & $3.5\times10^{-2}$ & 0.55 & 15.5 &  14.5 & 11.1 &  2.7 &  4.2 & 12.0\\ 
  Au-8 & 14.0 &  4.1 & $9.2\times10^{-3}$ & 0.79 &  9.8 &   9.4 & 28.6 &  4.5 &  4.6 & 13.1\\ 
  Au-9 & 53.6 &  5.6 & $4.2\times10^{-2}$ & 0.62 &  3.9 &   2.5 & 14.3 &  2.3 &  2.7 & 19.3\\ 
  Au-10 & 54.4 &  7.2 & $1.1\times10^{-1}$ & 0.43 & 12.5 &   5.1 & 23.3 &  2.3 &  3.0 & 25.6\\ 
  Au-11 & 109.2 &  7.2 & $1.9\times10^{-1}$ & 0.29 &  5.0 &   1.8 & 50.0 &  1.1 &  3.2 & 10.0\\ 
  Au-12 & 28.1 &  6.8 & $5.0\times10^{-2}$ & 0.79 & 13.5 &  10.5 &  9.1 &  2.9 &  3.5 & 16.9\\ 
  Au-13 & 151.5 &  8.8 & $1.3\times10^{-1}$ & 0.31 &  6.9 &   2.3 &  9.8 &  5.7 &  4.5 & 11.6\\ 
  Au-14 & 28.2 & 11.2 & $3.9\times10^{-2}$ & 0.67 & 15.5 &  19.6 & 13.7 &  4.7 &  6.5 & 40.9\\ 
  Au-15 & 15.5 &  6.0 & $2.3\times10^{-2}$ & 0.73 & 20.5 &  18.3 & 14.6 &  3.3 &  4.6 & 26.1\\ 
  Au-16 & 17.0 &  5.3 & $8.5\times10^{-3}$ & 0.81 & 13.0 &  10.4 & 84.3 &  3.4 &  3.5 & 14.7\\ 
  Au-17 & 63.4 & 10.1 & $5.9\times10^{-2}$ & 0.36 & 13.8 &   6.5 & 13.2 &  2.9 &  4.2 & 10.8\\ 
  Au-18 & 41.0 &  4.9 & $3.0\times10^{-2}$ & 0.68 &  4.8 &   2.7 & 25.4 &  2.6 &  3.6 & 11.2\\ 
  Au-19 & 30.2 &  5.4 & $3.0\times10^{-2}$ & 0.71 &  5.6 &   4.2 & 26.4 &  3.7 &  3.7 & 18.5\\ 
  Au-20 & 24.3 &  6.6 & $2.9\times10^{-2}$ & 0.49 & 19.2 &  11.2 & 38.5 &  4.3 &  5.0 & 12.2\\ 
  Au-21 & 26.4 &  7.0 & $2.7\times10^{-2}$ & 0.83 &  9.8 &  10.4 & 14.4 &  3.9 &  3.9 & 16.3\\ 
  Au-22 & 42.7 &  5.1 & $3.3\times10^{-2}$ & 0.51 & 14.7 &   4.9 & 13.0 &  2.5 &  2.0 & 14.2\\ 
  Au-23 & 41.8 &  8.0 & $4.4\times10^{-2}$ & 0.63 &  9.3 &   6.2 & 25.6 &  3.6 &  3.3 & 32.9\\ 
  Au-24 & 31.7 &  4.6 & $2.1\times10^{-2}$ & 0.63 &  7.8 &   3.7 & -747.4 &  3.3 &  3.4 & 14.2\\ 
  Au-25 &  9.0 &  4.7 & $1.1\times10^{-2}$ & 0.73 & 20.8 &  30.3 & 11.6 &  3.4 &  5.1 & 16.4\\ 
  Au-26 & 280.8 & 14.3 & $2.5\times10^{-1}$ & 0.46 &  7.0 &   2.0 & 15.4 &  3.0 &  2.4 & 22.4\\ 
  Au-27 & 40.3 &  6.4 & $3.1\times10^{-2}$ & 0.81 &  5.7 &   4.0 & 21.6 &  3.2 &  3.1 & 21.3\\ 
  Au-28 & 77.5 &  7.9 & $1.5\times10^{-1}$ & 0.74 & 11.9 &   4.1 & 15.7 &  1.9 &  1.8 & 18.4\\ 
  Au-29 & 57.2 & 10.2 & $6.6\times10^{-2}$ & 0.61 & 14.2 &   7.5 & 12.5 &  4.5 & 16.5 & 16.3\\ 
  Au-30 & 146.1 &  7.9 & $9.4\times10^{-2}$ & 0.37 &  7.0 &   2.1 & 12.7 &  4.5 &  3.5 & 12.3\\ 
  \hline\hline
  \end{tabular}
  \label{tab}
  \caption{Magnetic field properties of the Auriga galaxies at $z=0$. For general properties of the disks see \citet{Auriga}.}
\end{table*}

A detailed summary of the Auriga project can be found in \citet{Auriga}. In addition, previous analysis of the Auriga galaxies that focused on the stellar disk and stellar halo \citep{Gomez2016,Monachesi2016,Grand2016,Gomez2016b}, and the HI disk \citep{Marinacci2017}, has shown that the simulations reproduce remarkably well a wide range of present-day observables of Milky Way-like galaxies. In the present work, we now turn to an investigation of the magnetic field strength in these systems. Non-magnetic properties of the simulated galaxies that we refer to below are taken from \citet{Auriga}, if not stated otherwise. All distances quoted in this paper are in physical units.

\begin{figure*}
  \includegraphics[width=\textwidth]{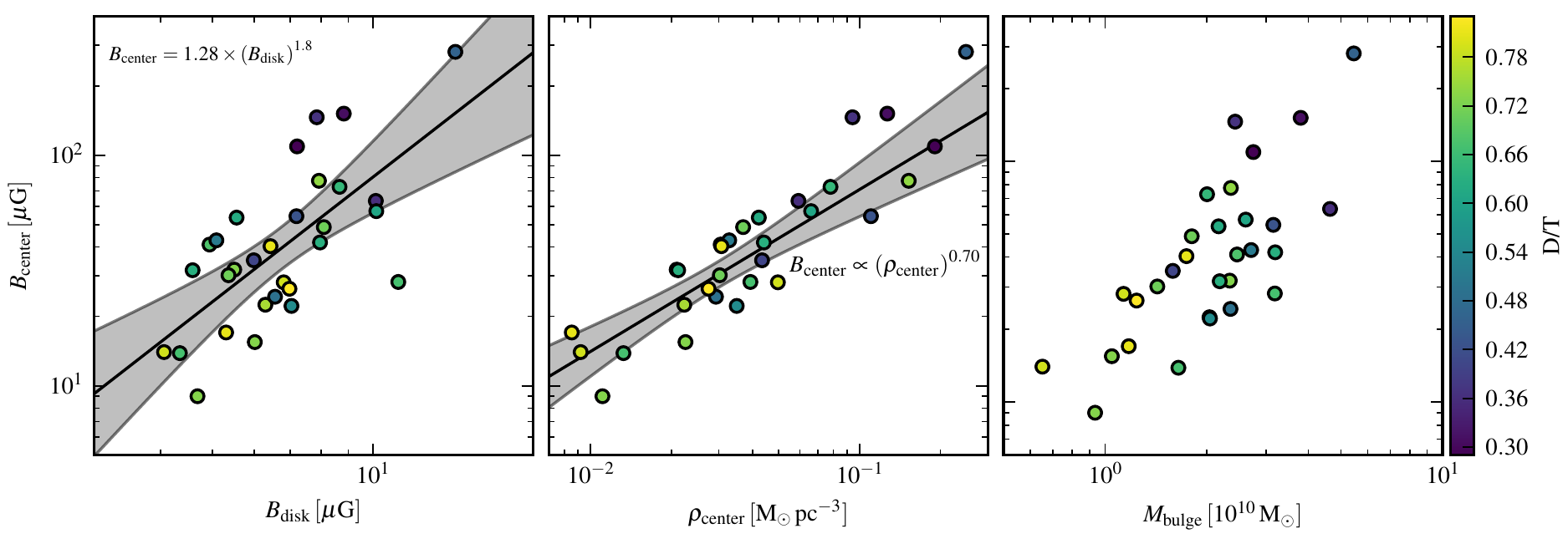}
  \caption{Correlation of the magnetic field strength in the galaxy center with the average magnetic field strength in the disk (left panel), with the central gas density (middle panel), and with the bulge mass (right panel), for all systems. The colour-coding represents the fraction of stellar mass in the disk compared to total stellar mass in the halo ($D/T$). The black line in the left panel shows the best fit relation between central and average disk magnetic field and the grey area the $95\%$ confidence interval. Similarly, in the central panel the black line shows the best fit relation with a power-law index of $0.70$. Here, we only included galaxies with $D/T>0.5$. If we include all galaxies the relation becomes slightly steeper and we find a best fit power-law index of $0.82$.}
  \label{fig:bstrength}
\end{figure*}

\section{The strength of the disk magnetic field at Z=0}

\label{sec:disk}

In Figure~\ref{fig:disks}, we show projections and profiles of the 6 halos that we have or will have higher resolution runs available. They were selected to have large disks (Au-16 and Au-24), to be a close equivalent to the Milky Way (Au-6), and for interesting satellite interactions (Au-21, Au-23, and Au-27). The projected magnetic field strength in every pixel of the images is calculated from the line-of-sight integral of the magnetic energy density, i.e.
\begin{equation}
  B\left(x,y\right) = \left(  \frac{1}{2h} \int_{-h}^{+h} B\left(x,y,z\right)^2 \rm{d}z \right)^{1/2}
\end{equation}
Similarly, the profiles show the magnetic field strength equivalent to the total magnetic energy in every ring for the radial profiles and in every cylinder for the vertical profiles, always centered on the disk. Here, the average magnetic field strength in a radial bin with a radial size of $1\,\rm{kpc}$ is computed by
\begin{equation}
  B\left(r\right) = \left(  \frac{ \int_{r-0.5\rm{kpc}}^{r+0.5\rm{kpc}} \int_{-1\rm{kpc}}^{+1\rm{kpc}}  \int_{0}^{2\pi} B\left(r,\phi,z\right)^2 r\, \rm{d}\phi\, \rm{d}z\, \rm{d}r }
  { \int_{r-0.5\rm{kpc}}^{r+0.5\rm{kpc}} \int_{-1\rm{kpc}}^{+1\rm{kpc}}  \int_{0}^{2\pi} r\, \rm{d}\phi\, \rm{d}z\, \rm{d}r }  \right)^{1/2}.
\end{equation}

The magnetic field strength in the center of the disks typically reaches a few $10\,\rm{\mu G}$ and drops to a few $\rm{\mu G}$ at larger radii ($10-30\,\rm{kpc}$). The radial profile of the magnetic field strength in the disk can be fit very well with two exponentials that describe the magnetic field strength in two regions separated by a break radius $r^B_0$, i.e.~a four parameter fit defined by
\begin{equation}
  B\left(r\right) = 
\begin{cases}
  B_\mathrm{center} \times e^{ \left[- r / r^B_\mathrm{inner} \right] } & \mbox{if} \ r < r^B_0 \\
  B_\mathrm{center} \times e^{ \left[ - r^B_0 / r^B_\mathrm{inner} - \left( r - r^B_0 \right) / r^B_\mathrm{outer} \right] } & \mbox{if} \ r \ge r^B_0.
\end{cases}
\end{equation}
Here, $B_\mathrm{center}$ is the magnetic field strength at $r=0$, $r^B_\mathrm{inner}$ and $r^B_\mathrm{outer}$ are the scale radii of the inner and outer exponentials, and $r^B_0$ is the radius where the slope changes from the inner to the outer exponential. For our set of halos, the inner scale radius is always smaller than the outer scale radius. The values obtained by a least squares fit in log-space, i.e. we fit a linear function to $\log{\left(B\left(r\right)\right)}$, for the inner scale radius typically range from $1\,\rm{kpc}$ to $10\,\rm{kpc}$, for the outer scale radius we obtain values between $10\,\rm{kpc}$ and $40\,\rm{kpc}$, and the break typically occurs at a radius between $5\,\rm{kpc}$ and $15\,\rm{kpc}$. Note that some of the halos (e.g.~Au~21) are well fitted with a single exponential and the break radius inferred from the fit becomes meaningless. There is no obvious one-to-one correlation between the radial profiles of the gas density and the magnetic field strength; commonly the gas density exhibits a more complex radial structure. Interestingly, there is significant azimuthal structure in the magnetic field strength in the disk, an analysis of which is deferred to a forthcoming paper.

The edge-on projections of the magnetic field strength for the disk galaxies in Fig.~\ref{fig:disks} show a large variety in the spatial extent of strong magnetic fields within the halo. At the extreme end, highly magnetized outflows (e.g.~Au 21) occur with field strengths of several $\mu G$, but configurations where the magnetic field strength drops significantly below $1\,\mu G$ by a height of $10\,\rm{kpc}$ (e.g. Au 16) can also be found. In general, the vertical profile of the magnetic field strength can also be described by a double exponential. The inner and outer vertical scale heights are typically in the range of $2\,\rm{kpc}$ to $5\,\rm{kpc}$, and $10\,\rm{kpc}$ to $20\,\rm{kpc}$, respectively, with a break at $2\,\rm{kpc}$ to $5\,\rm{kpc}$. For most galaxies, the vertical profile of the magnetic field strength is closely mirroring the vertical profile of the gas density, which also can be described well by a double exponential profile with the same break, but with a three times larger scale height. This strongly suggests that the vertical profile of the magnetic field strength is set by the galactic winds that are responsible for shaping the vertical gas density profile of the circumgalactic medium. We analyse the relation between the two vertical profiles in more detail below. For a list of the magnetic field properties of all galaxies at $z=0$ see table~\ref{tab}.

\subsection{Central magnetic field strength at $z=0$}

Estimates inferred from integrated radio continuum observations exist for the magnetic field strength at the very center of disk galaxies, the average field strength over the whole disk, and for radial profiles of the magnetic field strength for a small number of disk galaxies. Note, that except for the estimates of the magnetic field strength at the very center of the Milky Way, all these estimates assume equipartition between magnetic energy and the energy of the cosmic ray protons  and a fixed ratio between the energy density of cosmic ray protons and cosmic ray electrons that generate the radio synchrotron emission.

The central magnetic field strength in our galaxies at $z=0$ ranges from about $10\,\mu G$ up to $300\,\mu G$. However, the strongest magnetic fields with a strength above $100\,\mu G$ are only found in bulge-dominated galaxies. A central field strength of several $10\, \mu G$ as found in the disk-dominated galaxies of our sample is slightly smaller but possibly still consistent with observational estimates that report $50\, \mu G$ to $100\, \mu G$ on scales of $400\,\rm{pc}$ \citep{Crocker2010} for the magnetic field strength in the center of the Milky Way. These scales are resolved in the central regions of our simulations.

The average magnetic field strengths for the whole disk range between a few $\mu G$ to about $10\,\mu G$ across our samples. These strengths are at the lower end of the average magnetic field strength in nearby galaxies inferred from integrated radio continuum observations \citep[see, e.g.][]{Fletcher2010}, which cite values of $17 \pm 3 \, \mu G$ and assumes equipartition between magnetic energy density and the energy density of the cosmic ray electrons that generate the radio synchrotron emission. A possible explanation of this difference may lie in the reduced strength of small-scale turbulence in the ISM of our simulations, which is a direct consequence of our sub-resolution ISM model \citep{Springel2003} and is bound to reduce the turbulent component of the represented magnetic field. Indeed, the average field strength we find is more consistent with the estimated strength of the {\em regular}, large-scale component of the magnetic field of about $5 \pm 3 \, \mu G$ \citep{Fletcher2010}. However, we note that there are still significant uncertainties in the observational magnetic field strength estimates that preclude us from drawing any strong conclusions.

The central and average magnetic field strengths in our galaxies are correlated as shown in the left panel of Fig.~\ref{fig:bstrength}, but exhibit significant scatter. The correlation is at least partly expected as the strong magnetic fields in the small volume in the center of each galaxy still contribute significantly to the total magnetic energy in the disk.

\begin{figure}
  \centering
  \includegraphics[width=\linewidth]{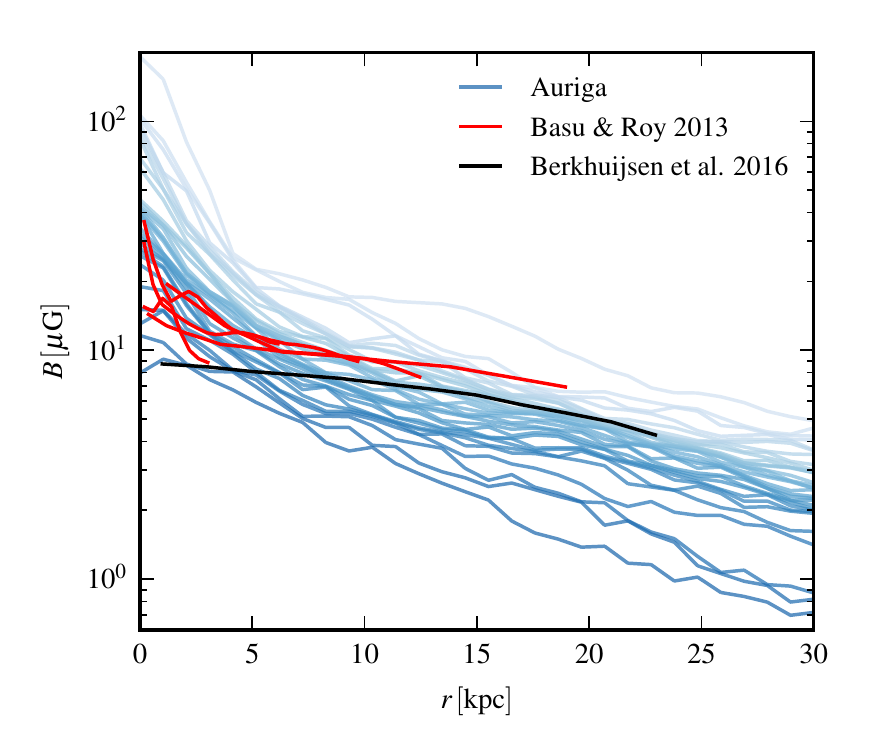}
  \caption{Radial profiles of the magnetic field strength at $z=0$, measured in rings with a height of $2\,\rm{kpc}$, for all halos. The blue lines show the profiles of the individual halos. The shading represents the order of central magnetic field strength. The red lines show the observed radial profiles of five nearby galaxies \citep{Basu2013}, the black line shows the radial profile for M101 \citep{Berkhuijsen2016}. All observed profiles assume equipartition between magnetic energy density and the energy density of cosmic ray protons  and a fixed ratio between the energy density of cosmic ray electrons and cosmic ray protons.}
  \label{fig:rprof}
\end{figure}

In the middle and right panels of Fig.~\ref{fig:bstrength}, we show the correlations between central magnetic field strength and central gas density and bulge mass, respectively. The central magnetic field strength tightly correlates with the gas density in the center. The relation exhibits a slope of $B \propto \rho^{0.7}$, close to the value of $B \propto \rho^{2/3}$ expected for isotropic adiabatic compression, with a small deviation for the most bulge-dominated galaxies with the strongest central magnetic fields. The central magnetic field strength in disk-dominated galaxies is therefore consistent with not being set by local dynamo amplification, but rather by adiabatic compression of the already magnetised gas in the disk when it flows to the center. Note that this requires that all the galaxies in our sample reach a similar magnetic field strength in the disk at a comparable gas density.

\begin{figure}
  \centering
  \includegraphics[width=\linewidth]{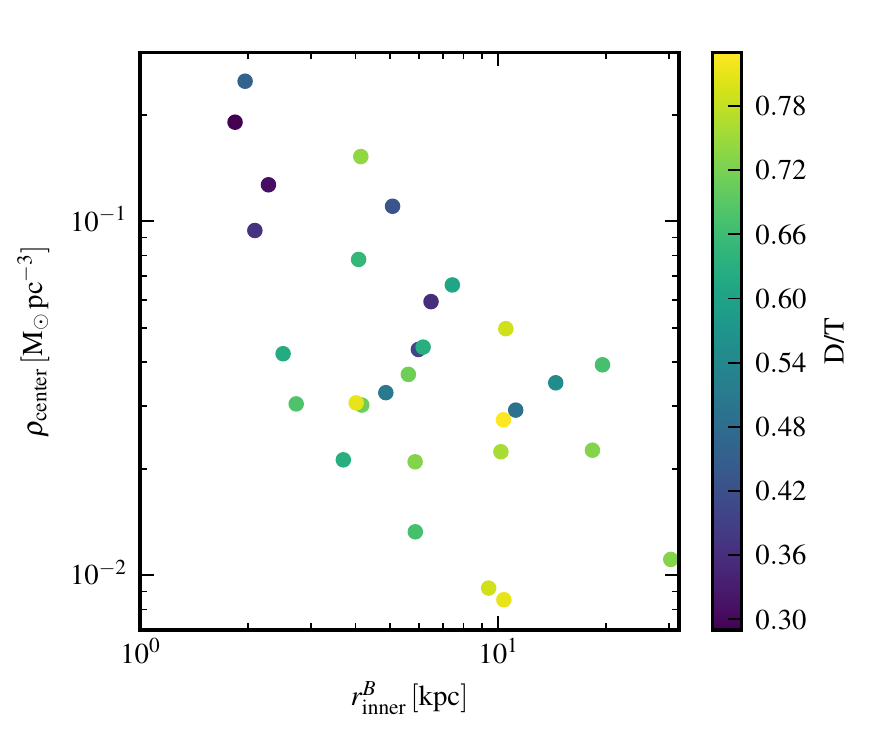}
  \includegraphics[width=\linewidth]{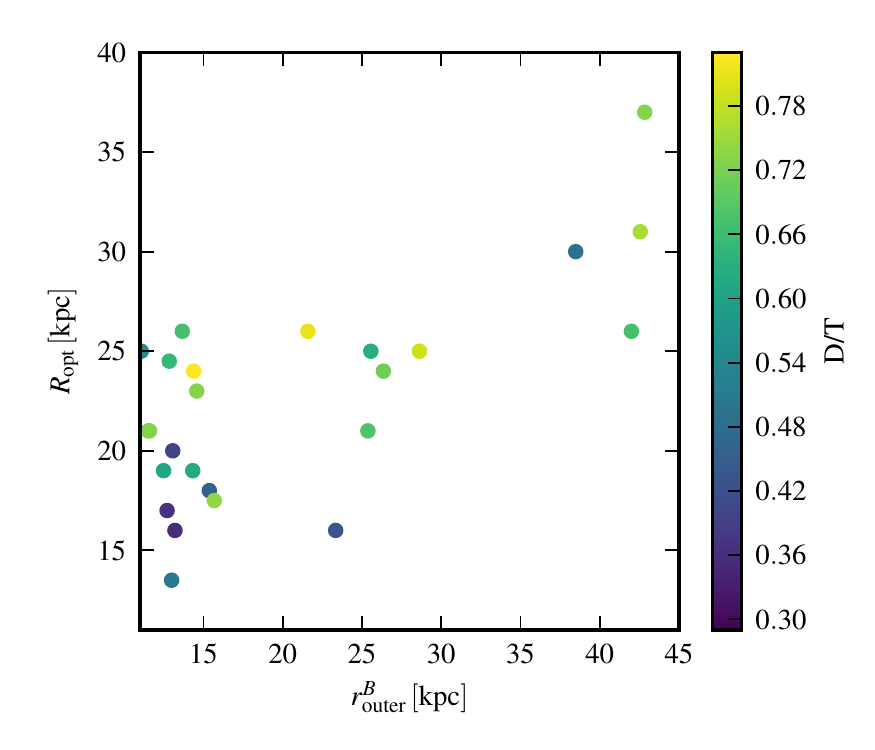}
  \caption{Inner scale radius versus central density (upper panel), and outer scale radius of the magnetic field strength versus optical radius of the stellar disk (lower panel). The colour-coding denotes the fraction of stellar mass in the disk compared to total stellar mass in the halo ($D/T$).}
  \label{fig:radialfits}
\end{figure}

The central magnetic field strength correlates more weakly with the mass of the bulge and the mass of the central black hole, and only slightly with the total stellar mass of the galaxy. We do not see a correlation between halo mass and magnetic field strength for our sample. This is mainly a result of the small range of halo masses (only a factor of $2$) and the sufficiently strong scatter, making the scatter in the magnetic field strength larger than any aspect of systematic variation.
These trends are again consistent with being driven primarily by the gas density in the center, which is higher for galaxies with a larger bulge mass and therefore deeper central gravitational potentials. The weak correlation with stellar mass is likely a secondary effect that can be explained by the weak correlation between bulge mass and stellar mass.

\begin{figure*}
  \centering
  \includegraphics[width=\textwidth]{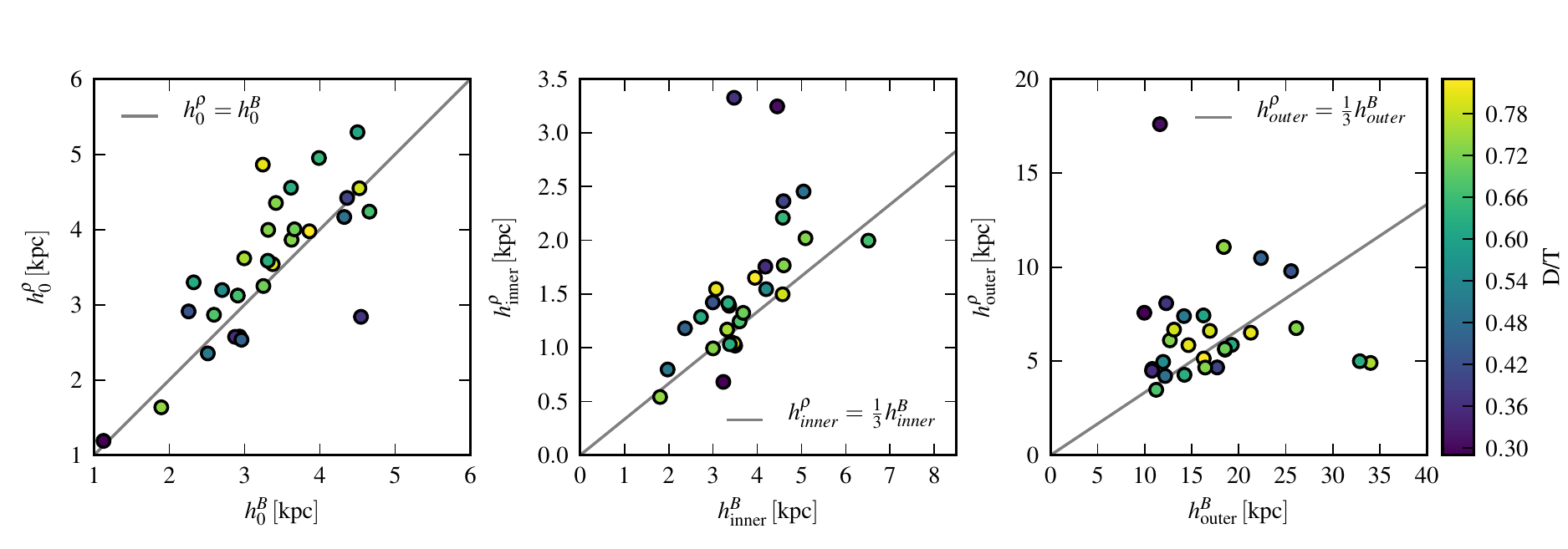}
  \caption{Characterisation of the vertical profile of the magnetic field strength. The left panel shows the vertical position of the break between inner and outer exponential for magnetic field strength $h_0^B$ and density $h_0^\rho$. The middle and right panel give the inner and outer scale height of the magnetic field strength versus the inner and outer scale height of the gas density. The grey line in the left panel shows equal heights of the break for magnetic field and density. The grey lines in the middle and right plot indicate a 3:1 ratio between the vertical scale heights of the magnetic field and the density.}
  \label{fig:verticalfits}
\end{figure*}

\subsection{Radial magnetic field profiles at $z=0$}

The radial magnetic field profiles in the disk at $z=0$ are shown in Fig.~\ref{fig:rprof} compared to observational data. At all radii, the magnetic field strength varies by about an order of magnitude between the different halos, although $66\%$ of the halos only span a range of about a factor of two, except for the center of the galaxies where the range is larger. The ordering of the profile is not strictly monotonic, i.e.~the disk with the weakest magnetic field in the center does not have the weakest magnetic field everywhere. Nevertheless, galaxies with higher average magnetic field strength tend to have higher magnetic field strengths at all radii, consistent with our previous result that the central magnetic field strength correlates with the average field strength over the whole disk.

The profiles in our simulated galaxies agree well with observed galaxies, but more and better data and in particular a self-consistent modelling of the cosmic ray electron population and its radio emission are needed to better understand the diversity of magnetic fields in galaxies.

The inner scale radii of the magnetic field strength correlate with the central densities; higher central gas densities come with smaller inner scale radii and therefore a steeper decline of the magnetic field strength, as shown in Fig.~\ref{fig:radialfits}, reinforcing the idea that the central gas density sets the magnetic field strength in the inner part of the galaxy. The outer scale radius is correlated with the optical radius of the galaxy, indicating that it is set by the global, large-scale properties of the galaxy. The position of the radial break that separates the inner and the outer exponential does not seem to correlate with other galaxy quantities in any obvious way.

\subsection{Vertical magnetic field profiles at $z=0$}

As shown for a subset of 6 galaxies in Fig.~\ref{fig:disks}, the vertical structure of the magnetic field strength as well as the density profile can be very well described by two joint exponential profiles. Fig.~\ref{fig:verticalfits} shows the correlation between the fit parameters obtained for the magnetic field and the gas density. The inner exponential transitions into the outer exponential at a height between $2\,\rm{kpc}$ and $5\,\rm{kpc}$, and this break is essentially at the same place for the gas density ($h^\rho_0$) and the magnetic field strength ($h^B_0$).

\begin{figure*}
  \centering
  \includegraphics[width=\textwidth]{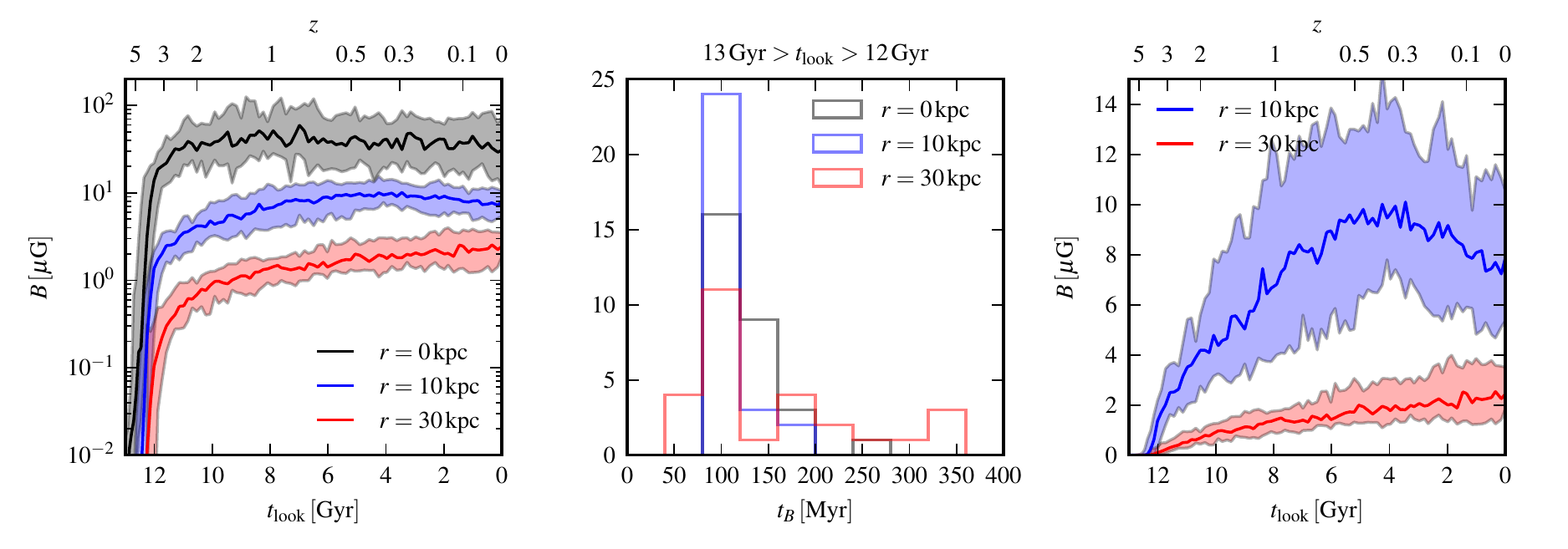}
  \caption{Magnetic field amplification in the disk. The left panel shows the median magnetic field strength for all halos in a radius of $1\,\rm{kpc}$ (grey) and in rings with a radius of $10\,\rm{kpc}$ (blue) and $30\,\rm{kpc}$ (red), a radial extent of $1\,\rm{kpc}$, and a height of $1\,\rm{kpc}$. The bands represent $66\%$ of the halos. The middle panel shows a histogram of the amplification timescales at lookback times larger than $12\,\rm{Gyr}$. The right panel shows the median magnetic field strength for all halos for rings at $10\,\rm{kpc}$ (blue) and $30\,\rm{kpc}$ (red), respectively, on a linear scale.}
  \label{fig:bamp}
\end{figure*}

The inner scale heights of gas density and magnetic field strength in the plane of the gas disk are strongly correlated. The inner magnetic scale height is roughly three times as large as the scale height of the gas density, which translates to $B \propto \rho^{1/3}$. This scaling can be interpreted as being intermediate between one-dimensional compression/expansion along magnetic field lines (which leads to $B \propto \rho^{0} = \rm{const}$), isotropic compression and expansion (which leads to the usual scaling $B \propto \rho^{2/3}$), and finally expansion/compression only perpendicular to magnetic field lines (which leads to $B \propto \rho$). Therefore, the vertical scaling of $B \propto \rho^{1/3}$ we observe argues that compression and expansion of gas occur predominantly along magnetic field lines, with a much smaller but non-negligible contribution perpendicular to the magnetic field lines. Moreover, this also implies that the magnetic field above and below the disk is predominantly turned into the vertical direction by the galactic wind. We will investigate this process in more detail in our companion paper.

The outer scale heights of the magnetic field strength and gas density are also consistent with $B \propto \rho^{1/3}$, but with significantly larger scatter. Moreover, the outer scale height is typically much larger than the inner scale height, with typical values of $10\,\rm{kpc}$ to $30\,\rm{kpc}$ for the outer magnetic scale height. Both, the outer magnetic and gas scale heights, are likely to be strongly influenced by our model for galactic winds \citep{Auriga,Marinacci2014}. For the outer scale height, we obtain a consistent picture in which the velocity field is dominated by vertical outflows and the magnetic field is dominated by its vertical component as well. Taken together, this then causes the observed scaling. The origin of the scaling in the disk is less obvious, although magnetic field and velocity field are both dominated by the azimuthal component. Note that the only two galaxies that do not follow this relation are strongly bulge-dominated and therefore lack highly ordered velocity and magnetic fields on large scales.

\begin{figure}
  \centering
  \includegraphics[width=\linewidth]{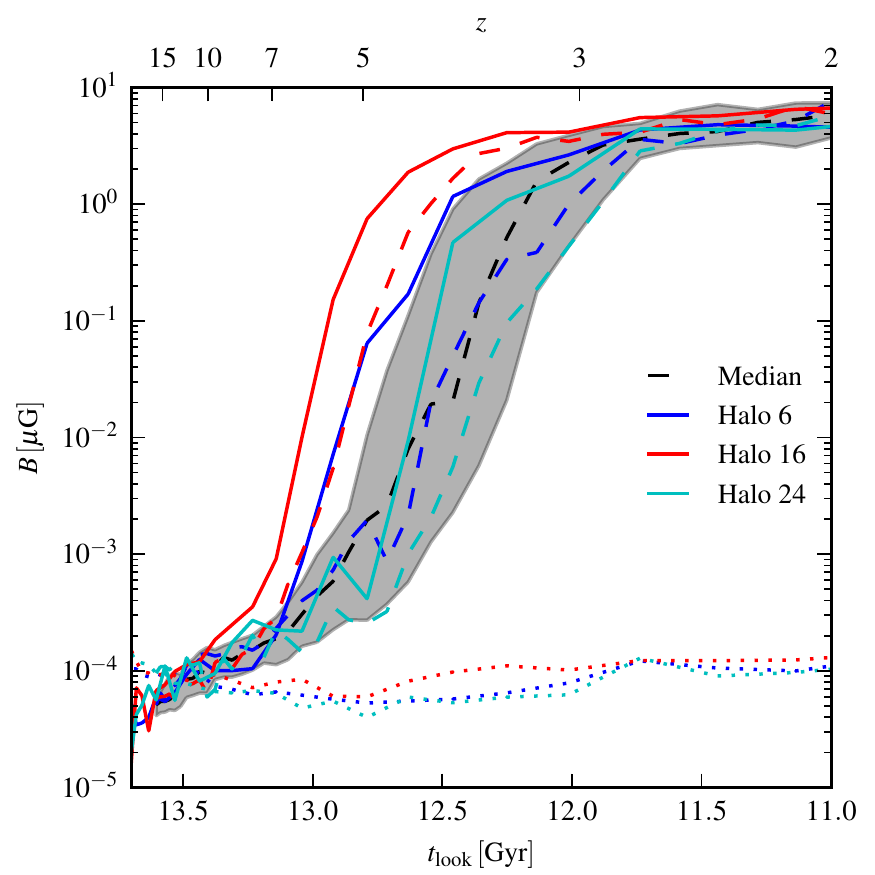}
  \caption{Magnetic field amplification at high redshift computed as the volume weighted rms magnetic field strength in a sphere with a radius of $10\,\mathrm{kpc}$ centered on the potential minimum of the halo. Solid lines show the magnetic field strength for halo 6, halo 16, and halo 24 at level 3 resolution. Dashed lines show the same halos at level 4 (standard) resolution. Dotted lines show the theoretical magnetic field strength for the level 3 halos in case of flux freezing (i.e. $B\propto \rho^{2/3}$). The dashed black line denotes the median magnetic field strength for all level 4 halos, and the grey area encloses $66\%$ of their distribution.}
  \label{fig:bampearly}
\end{figure}

\begin{figure*}
  \centering
  \includegraphics[width=\textwidth]{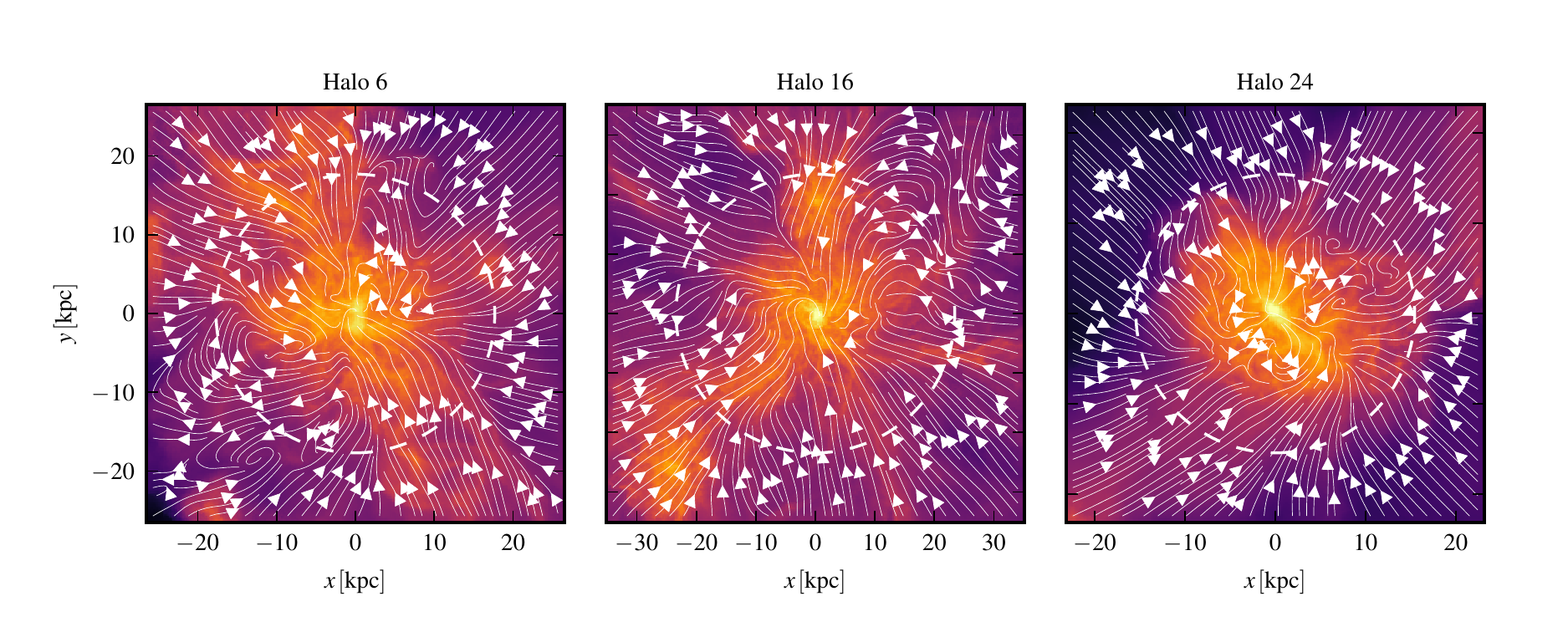}
  \caption{Gas density projections of three halos at level 3 resolution at $z=5$. The colorcoding shows the logarithmic column density of the gas. The streamlines show the direction of the projected velocity field. The dashed line describes the virial radius ($r_{200,c}$) of the halos at this time}
  \label{fig:vel}
\end{figure*}

\section{Magnetic field amplification}

\label{sec:amplification}

To reach field strengths of order $\mu G$ or more in the galaxies at $z=0$, the magnetic field has to be amplified by many orders of magnitude from the small, uniform primordial seed field we adopted in the initial conditions. The required amplification is much larger than the amplification obtained by adiabatic contraction when the gas is accreted on galaxies. Therefore, an efficient dynamo is required to increase the magnetic field strength to the values found in the simulation.

There are two main types of dynamos that have been claimed to be active in galaxies, a turbulent small-scale dynamo and an $\alpha$-$\Omega$-dynamo. Both dynamos can in principle exhibit exponential as well as linear growth of the magnetic energy.

The turbulent dynamo operates in a turbulent medium. As long as the magnetic pressure is irrelevant, i.e. the dynamo operates in the kinetic regime, it will exponentially amplify the magnetic energy, and it will do so on the smallest scales first. Once equipartition is reached on the smallest scales, magnetic energy can be moved to larger scales by an inverse cascade. In contrast to the kinetic phase, this phase only leads to a linear amplification of the magnetic energy \citep[see, e.g.][]{Federrath2016}. The turbulent dynamo characteristically leads to a chaotic, small-scale magnetic field that carries the same energy on each spatial component.

The $\alpha$-$\Omega$-dynamo instead exploits the differential rotation in a disk ($\Omega$-effect) in combination with small-scale vertical motions ($\alpha$-effect). In its simplified form as galactic mean-field dynamo it can lead to exponential or linear growth of the magnetic energy depending on the relative size of the vertical and rotational terms \citep{Shukurov2006}. It characteristically generates a large-scale ordered magnetic field that is dominated by its azimuthal component.

\begin{figure*}
  \centering
  \includegraphics[width=\textwidth]{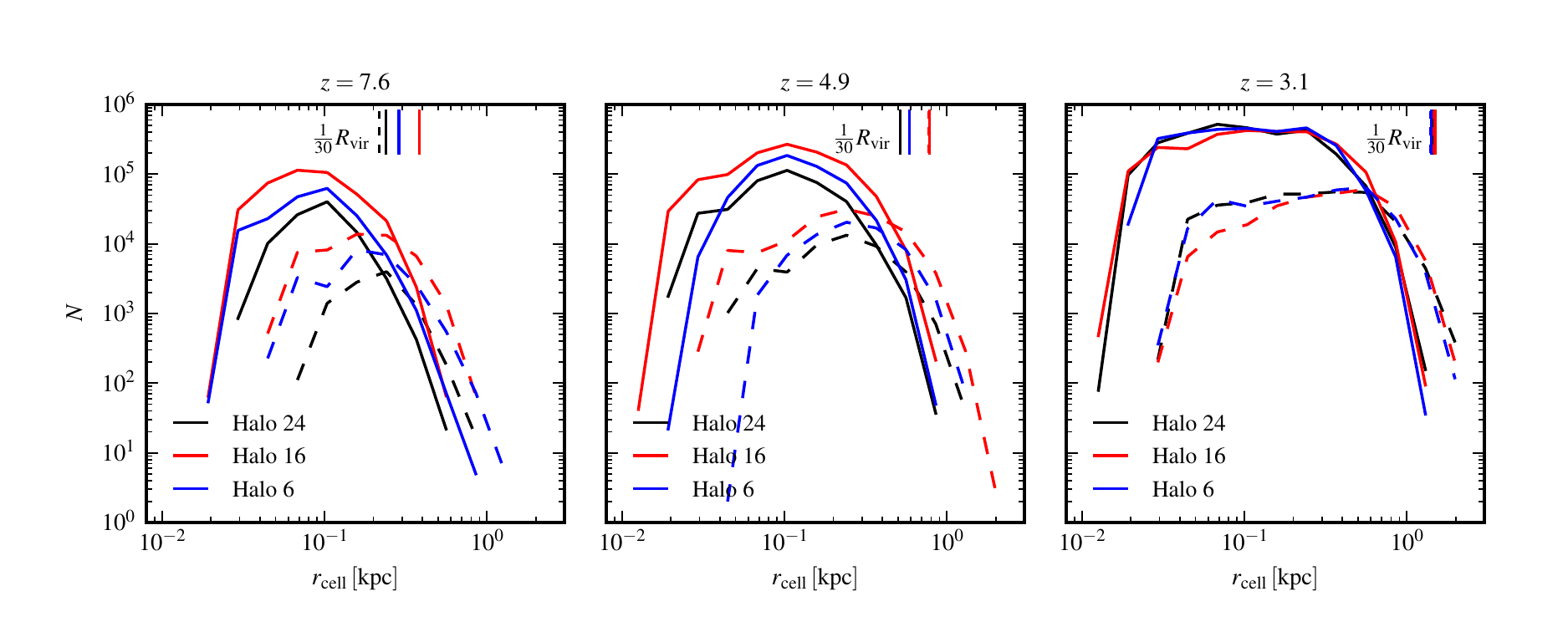}
  \caption{Histogram of cell radii for 3 halos at redshifts between $z=7.6$ (left panel) to $z=3.1$ (right panel). Solid lines show the high resolution level 3 runs, dashed lines show the standard level 4 runs. Vertical solid lines denote 1/30 of $r_{200,\mathrm{c}}$ of the halo.}
  \label{fig:res_turb}
\end{figure*}

\begin{figure*}
  \centering
  \includegraphics[width=\textwidth]{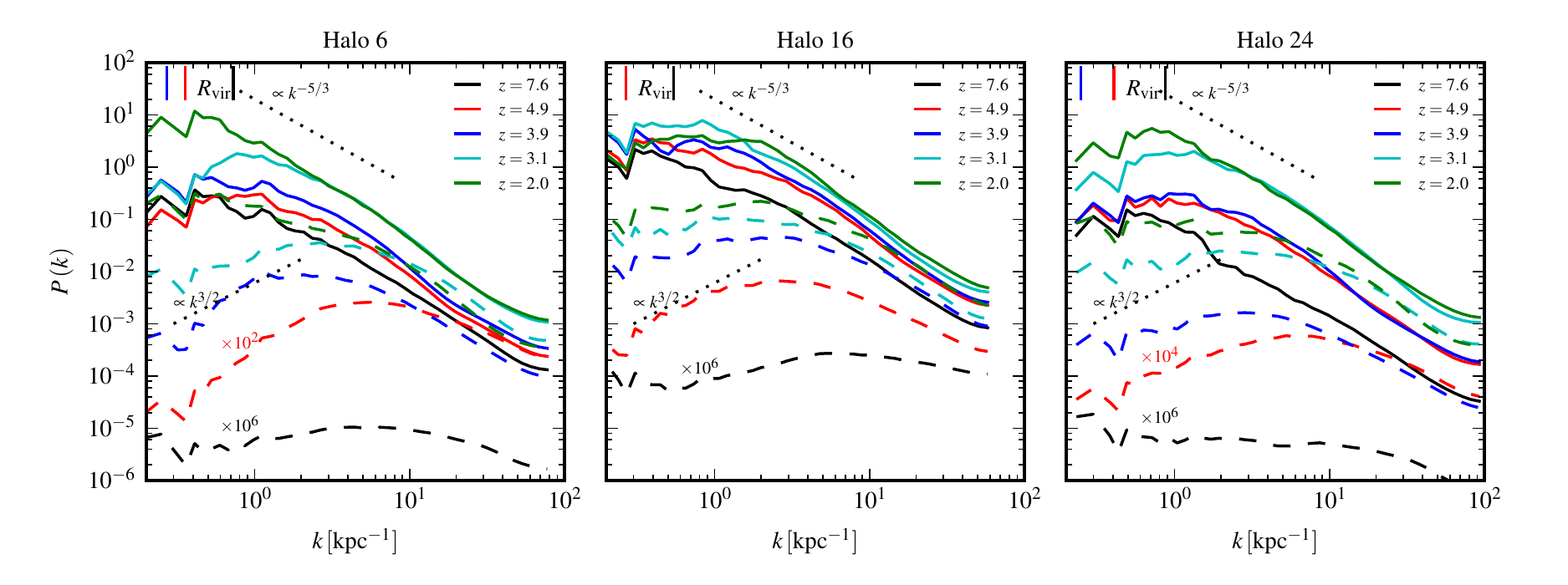}
  \caption{Power spectra of kinetic energy (solid lines) and magnetic energy (dashed lines) for 3 halos at level 3 resolution for different redshifts from $z=7.6$ to $z=2$.  The power spectra are computed using a zero-padded fast Fourier transform of the components of $\sqrt{\rho}\mathbfit{v}$ and $\mathbfit{B}$ for the gas within a constant physical radius. The radius was chosen for each halo individually as the corresponding virial radius ($r_{200,c}$) at $z=7.6$. The radii are $9\,\mathrm{kpc}$ for halo 6, $11\,\mathrm{kpc}$ for halo 16, and $7\,\mathrm{kpc}$ for halo 24. The black dotted lines show slopes $P\left(k\right) \propto k^{-5/3}$ of a Kolmogorov spectrum \citep{Kolmogorov1941}, and $P\left(k\right) \propto k^{3/2}$ of a Kazantsev spectrum \citep{Kazantsev1985} for comparison. Vertical lines in the top left of the panels show the virial radius. At lower redshift the virial is larger than the range of the plot.}
  \label{fig:ps}
\end{figure*}

Previously, we argued based on a single galaxy simulation that the magnetic field is first amplified exponentially by a small-scale dynamo and undergoes a second phase of linear growth owing to a galactic mean-field dynamo once the disk forms \citep{Pakmor2014}. Here, we can now reevaluate this conclusion based on the large sample of the Auriga simulations.

\subsection{The early exponential amplification phase}

The time evolution of the median magnetic field strength of all halos in the galactic centers and in rings with radii of $10\,\rm{kpc}$ and $30\,\rm{kpc}$, respectively, is shown in Fig.~\ref{fig:bamp}. Similar to previous work that focused either on the formation of a single galaxy \citep{Pakmor2013,Pakmor2014,Pakmor2016b} or on larger scales of whole halos \citep{Dolag1999,Beck2012,Marinacci2015}, we see an initial period of fast exponential amplification at all three radii. This period is shown in detail in Fig.~\ref{fig:bampearly}. At $z=10$, the magnetic field strength in the halos is identical to the value expected from flux freezing. This changes around $z=7$, when fast dynamo sets in. This process exponentially amplifies the magnetic field until it saturates around $z=3$. The time when saturation is reached varies between halos. For halos in which the dynamo sets in earlier, the magnetic field strength saturates earlier. Not also that the dynamo is only active in collapsed structures. Outside them the magnetic field strength evolves as expected from flux freezing \citep{Marinacci2015}.

For the high resolution level 3 runs the exponential amplification sets in earlier than for the standard resolution level 4 runs, and the amplification timescale is shorter as well. Both effects point to a turbulent magnetic dynamo that only occurs once the halo is sufficiently resolved. The velocity field in the halos at high redshift that is shown in Fig.~\ref{fig:vel} suggests a turbulent dynamo with an injection scale of roughly $L = r_{200,c}$. In this picture, when the halos form they initially are not sufficiently resolved to start the dynamo. However, they quickly grow in size as they accrete mass which increases the injection scale. Eventually the injection scale is sufficiently resolved and the turbulent dynamo starts. This interpretation is supported by the distribution of cell sizes in the halo that is shown in Fig.~\ref{fig:res_turb}. At $z=7.5$, only the high resolution run of Au~16 resolves the injection scale with more than 30 cells. For this halo the dynamo sets in first and saturates first. At $z=5$ all halos resolve the injection scale with more than 30 cells. The typical Jeans length for cells in the center of the halo at this time is $1\,\mathrm{kpc}$ with only a weak dependence on density owing to our equation of state for star-forming gas. This means that at this time we typically resolve the Jeans length by more than 32 cells in the center of the halo \citep{Schleicher2013}.

The most direct evidence for a turbulent small-scale dynamo comes from power spectra. Fig.~\ref{fig:ps} shows the evolution of power spectra of kinetic and magnetic energy in three high resolution level 3 halos from $z=7.6$ to $z=2$. We only show power spectra of the high resolution runs, but as discussed in Sec.~\ref{sec:resolution}, the evolution of the magnetic field strength is qualitatively consistent between the standard resolution and high resolution runs.

The shape of the kinetic energy power spectrum is consistent with a Kolmogorov spectrum \citep{Kolmogorov1941} for about an order of magnitude in $k$. Its amplitude increases with time as the halo grows until it converges around $z=3$. The magnetic energy power spectrum is essentially flat at $z=7.6$ before the dynamo sets in. After the dynamo starts, it transforms to a spectrum that is roughly consistent with a Kazantsev spectrum \citep{Kazantsev1985} on large scales.
\begin{figure*}
  \centering
  \includegraphics[width=\textwidth]{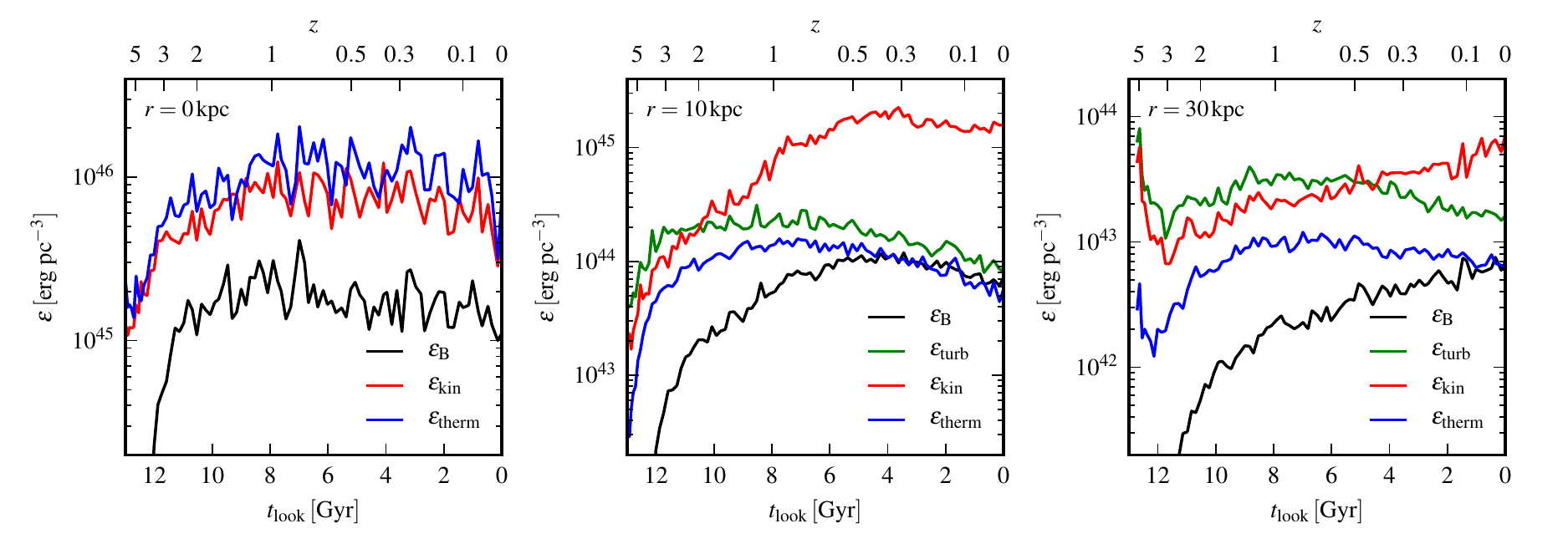}
  \caption{Time evolution of the median energy densities of magnetic energy, thermal energy, and kinetic energy in the disk for all galaxies. The panels show the median of the average energy densities at the center (left panel), and in rings at $10\,\rm{kpc}$ (middle panel) and $30\,\rm{kpc}$ (right panel). The rings have a radial extent of $1\,\rm{kpc}$, and a height of $1\,\rm{kpc}$. The kinetic energy is calculated in the rest frame of the halo. The turbulent energy is computed by substracting the local Keplerian velocity of a cell from the velocity of the cell. In the center there is no significant rotational support, so that $\epsilon_\mathrm{kin} \approx \epsilon_\mathrm{turb}$.}
   \label{fig:energy}
\end{figure*}

\begin{figure*}
  \centering
  \includegraphics[width=\textwidth]{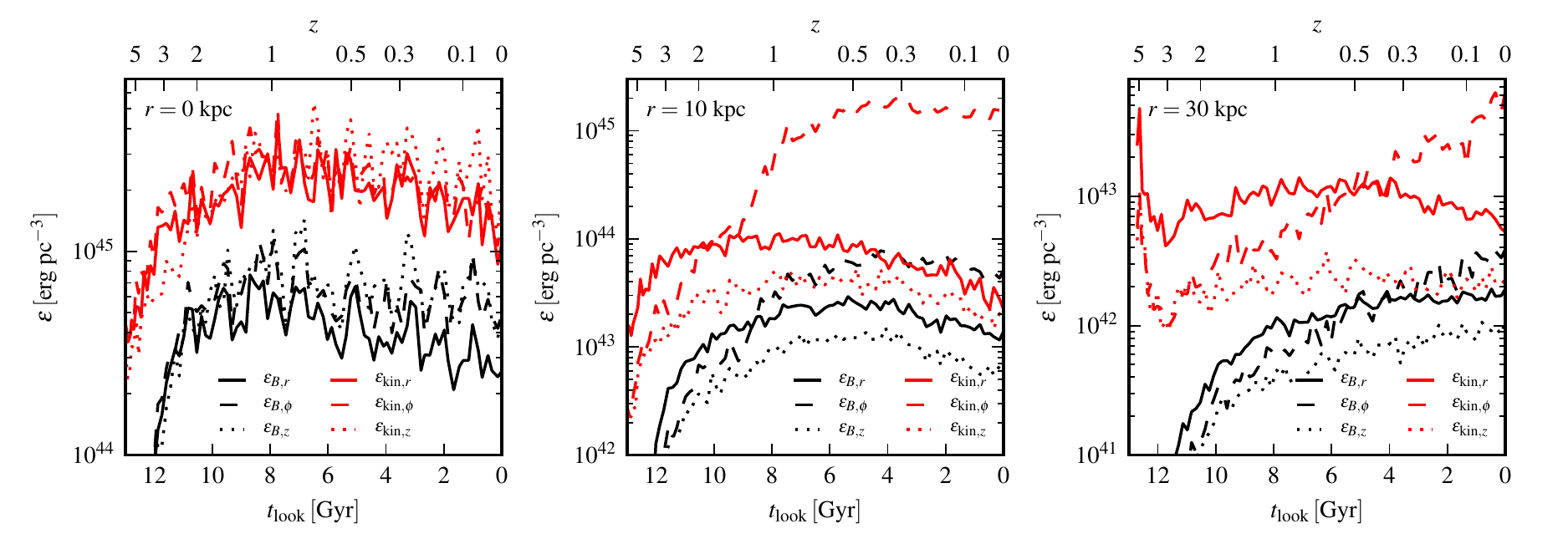}
  \caption{Time evolution of the median of the three components (radial, azimuthal, and vertical) of magnetic and kinetic energy density in the disk for all galaxies. The panels show the median of the average energy densities of the three components at the center (left panel), and in rings at $10\,\rm{kpc}$ (middle panel) and $30\,\rm{kpc}$ (right panel). The rings have a radial extent of $1\,\rm{kpc}$, and a height of $1\,\rm{kpc}$. The kinetic energy is calculated in the rest frame of the halo.}
   \label{fig:energydirs}
\end{figure*}

Note that both the time at which the turbulent dynamo sets in and the exponential timescale of the amplification are limited by the smallest scales we can resolve. Because we expect physical viscosity and resistivity to become important only on much smaller scales than those resolved in our simulation we likely significantly underestimate both, i.e. in reality the dynamo will set in earlier and amplify the magnetic field faster than in our simulation. However, the saturation field strength likely does not depend on resolution, therefore the evolution after we reach saturation will be the same. In addition, our ISM model does not include explicit local feedback from SN~II which may change the velocity field on small scales and influence the magnetic field amplification. Note however, that there are observational hints that the ISM turbulence is driven by gravity rather than local feedback for rapidly star-forming galaxies \citep{Krumholz2016}. Future simulations with a more sophisticated ISM model will be needed to investigate this in more detail.

Finally, the velocity field shown in Fig.~\ref{fig:vel} indicates that the turbulent velocity field is driven mostly by incoming filaments which suggests that the driving is dominated by solenoidal modes and that compressive modes are subdominant because the filaments generally do not align.

The e-folding timescale of the initial fast growth, obtained from fitting an exponential to the evolution of the magnetic field strength between $t_\mathrm{look}=13\,\rm{Gyr}$ and  $t_\mathrm{look}=12\,\rm{Gyr}$, is about $100\,\rm{Myr}$, as shown in the middle panel of Fig.~\ref{fig:bamp}. This is equivalent to a growth rate of $\Gamma \approx 10^{-2} \, \mathrm{Myr}^{-1}$. Our measured growth rate is about $10^4$ times slower than the analytical estimate for an accretion-driven, small-scale dynamo in a spherical galaxy \citep{Schober2013}, given by
\begin{equation}
  \Gamma \approx \frac{\varv\left(l\right)}{l} \mathrm{Re}^c,
\end{equation}
where $\varv\left(l\right)$ is the typical velocity on the scale of a typical turbulent eddy of size $l$, $\mathrm{Re}$ is the Reynolds number, and $c$ is a constant equal to $0.5$ for Kolmogorov turbulence. For the standard resolution runs the cells have a typical diameter of $500\,\mathrm{pc}$ (see Fig.~\ref{fig:res_turb}) and a velocity dispersion on this scale of $30\,\mathrm{km\,s^{-1}}$. Compared to the assumptions made by \citet{Schober2013} our ratio $\varv/l$ is about two orders of magnitude larger. However, \citet{Schober2013} assume a Reynolds number of $\mathrm{Re}=10^{11}$, many orders of magnitude larger than the effective Reynolds number in our simulation that is set by numerical viscosity. For a conservative estimate of $Re=10$ in our simulation we obtain an estimate for the growth rate of $\Gamma=0.1\,\mathrm{Myr}^{-1}$ equivalent to an amplification timescale of $10\,\mathrm{Myr}$. This estimated timescale is about a factor of $10$ faster than the measured timescale in the simulation. This may be caused by the rather large range in time we fit that is not limited to the pure kinetic dynamo and therefore likely underestimates the actual growth rate of the turbulent dynamo in the kinetic phase. In addition the loss of magnetic energy by star formation may also increase the effective growth time.

Note also, that the different components of the kinetic and magnetic energies are of similar strength at high redshift with a slight preference for the radial components in the outer parts of the halo (see also Fig.~\ref{fig:vel}).

Dynamos convert kinetic energy into magnetic energy. Once a dynamo is operating it will only end its amplification of an initially small magnetic field if the velocity field changes. This naturally happens when the magnetic pressure becomes large enough to have an effect on the velocity field. In this case, the magnetic field strength will necessarily be close to equipartition with the locally accessible kinetic energy. Alternatively, the magnetic field can saturate when loss terms reduce the magnetic energy at a rate compensating the amplification rate by the dynamo. In our simulations, loss terms for magnetic energy are star formation, which locks up the magnetic field of the gas that forms a star particle, and numerical resistivity on the grid scale. In addition, outflows can transport magnetic energy from the disk into the halo and act as a sink of magnetic energy for the galaxy.

As shown in Fig.~\ref{fig:ps} in the initial phase of exponential amplification the magnetic field saturates on the smallest scales first when the magnetic energy on these scales reaches $\approx 20\%$ of the kinetic energy. At this point, only larger modes continue to grow until they reach saturation as well. The evolution of the magnetic power spectrum is consistent with turbulent box simulations \citep{Federrath2016}. The turbulent dynamo saturates long before a disk forms and the velocity field in the galaxy becomes dominated by ordered large-scale rotation \citep{Pakmor2014}. The saturation at $\epsilon_B / \epsilon_{\rm{turb}} \approx 0.2$ is at the level expected from MHD turbulence simulations of the ISM for a range of realistic Reynolds and Prantl numbers \citep{Schekochihin2004,Cho2009,Kim2015,Federrath2016}. Note, that a more realistic smaller seed field will delay the time at which saturation is reached. In contrast, a more realistic Reynolds number will significantly increase the growth rate and make the dynamo reach saturation earlier. For realistic assumptions for both it is likely that saturation will be reached even earlier than in our simulation \citep{Schober2013}.

Finally, increasing the spatial resolution leads to a faster amplification of the magnetic field, but does not change the saturation level as shown in Sec.~\ref{sec:resolution}. We thus conclude that the initial growth of the magnetic field by a turbulent small-scale dynamo produces a completely unstructured small-scale magnetic field with a well-defined field strength.

\subsection{Linear amplification and the galactic dynamo}

After the initial exponential growth, the magnetic field strength at the center of the galaxy then stays constant until $z=0$. At larger radii the magnetic field strength keeps growing, but now only linearly with time. At a radius of $10\,\rm{kpc}$, the median magnetic field strength saturates roughly at $z=0.5$ and decreases slightly afterwards. Finally, at a radius of $30\,\rm{kpc}$, the magnetic field strength keeps rising until $z=0$.

In the center of the galaxy there is no significant contribution from rotation, and the kinetic energy can be used directly as a measure of the turbulent kinetic energy. At larger radii, a disk usually forms between $z=2$ and $z=1$. After this time, the kinetic energy is dominated by the ordered rotation of the disk. Therefore, we introduce an explicit estimate of the local turbulent energy density $\epsilon_\mathrm{turb}$ by subtracting the Kelperian velocity at every radius $r$, computed for simplicity from the total enclosed mass at this radius $M_\mathrm{enclosed} \left(r\right)$ as
\begin{eqnarray}
v_\mathrm{kep}\left(r\right) &=& \sqrt{ \frac{ G \, M_\mathrm{enclosed} \left(r\right) }{ r } } \\
\epsilon_\mathrm{turb} &=& \frac{ \rho }{ 2 } \left( \mathbfit{v} - v_\mathrm{kep}\left(r\right) \mathbfit{e}_\phi \right)^2,
\end{eqnarray}
where $\mathbfit{e}_\phi$ is the azimuthal unit vector.

\begin{figure*}
  \centering
  \includegraphics[width=\textwidth]{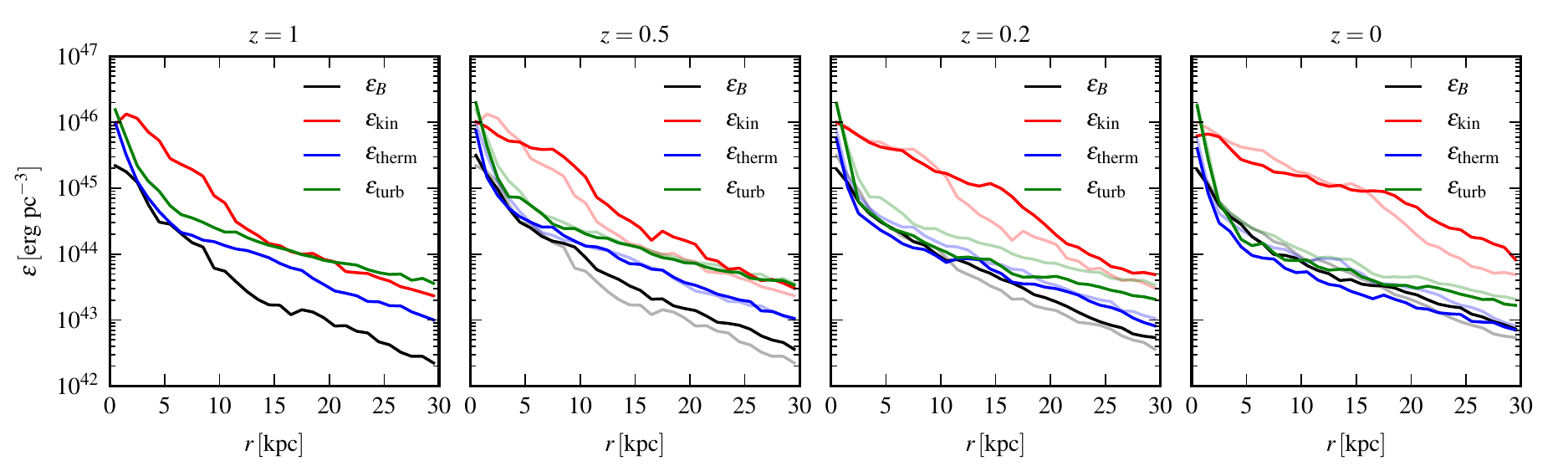}
  \caption{Median radial profiles of the different energy densities in the disk. The median is computed from all halos for every radial bin. The panels show the profiles at different times. The thin lines starting in the second panel show the profiles in the previous panel. The kinetic energy is calculated in the rest frame of the halo. The turbulent energy is computed by subtracting the local Keplerian velocity of a cell from its systemic velocity.}
  \label{fig:medianrprof}
\end{figure*}

\begin{figure*}
  \centering
  \includegraphics[width=\textwidth]{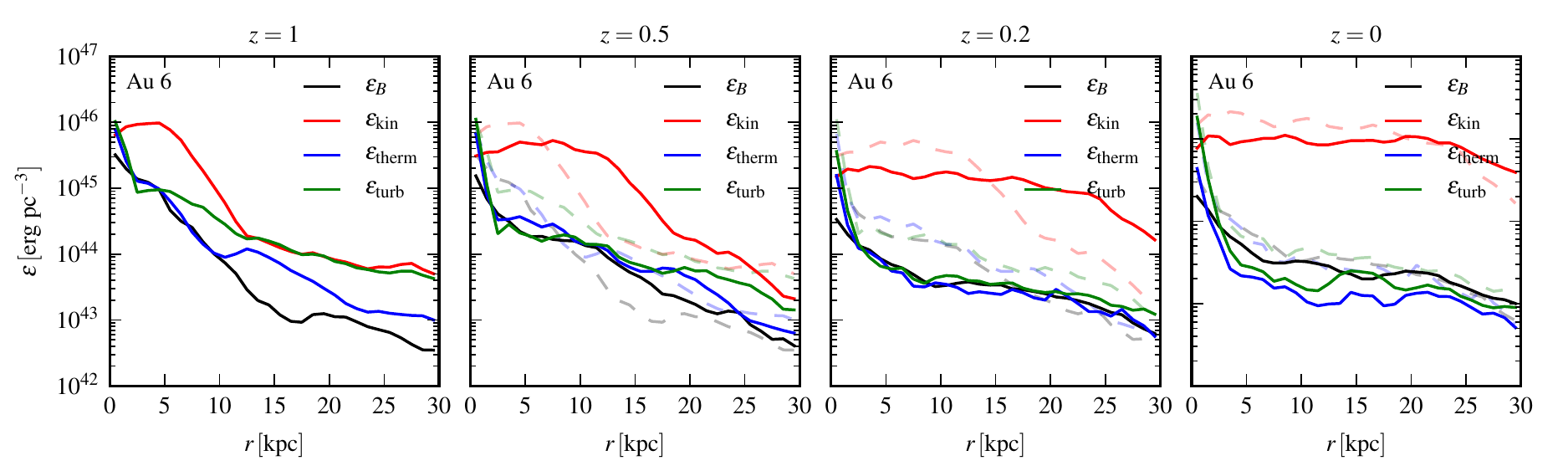}
  \includegraphics[width=\textwidth]{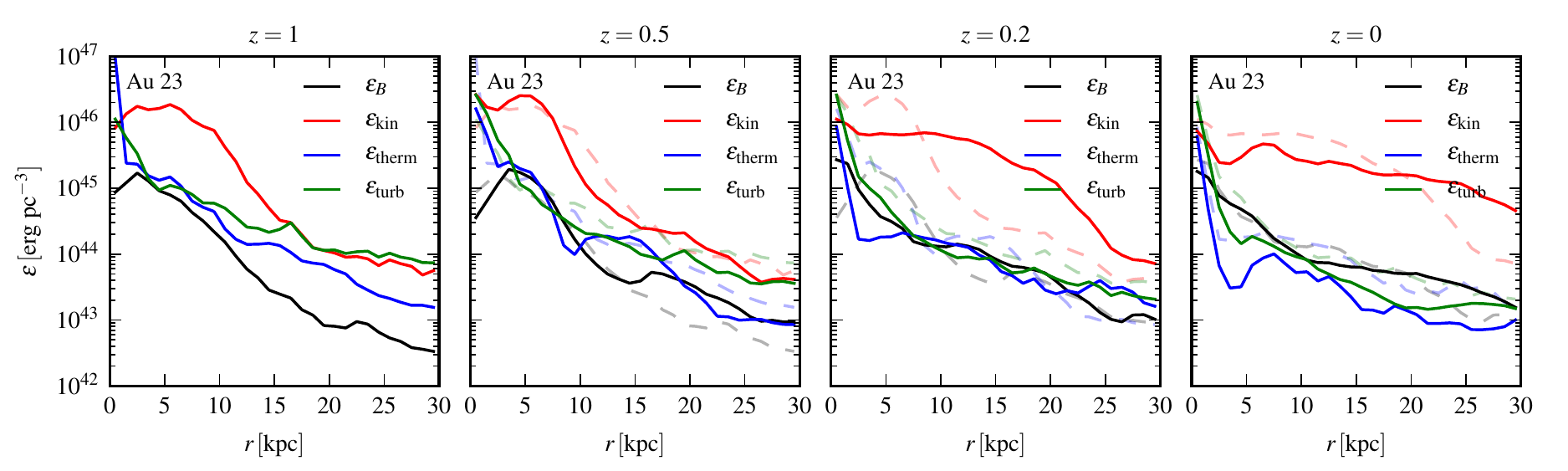}
  \caption{Radial profiles of the different energy densities in the disk for halo 6 (top tow) and halo 23 (bottom row). The panels display the profiles at different times. The thin dashed lines starting in the second panel show the profiles in the previous panel. The kinetic energy is calculated in the rest frame of the halo. The turbulent energy is computed by subtracting the local Keplerian velocity from the systemic velocity of a cell.}
   \label{fig:halosrprof}
\end{figure*}

In the center of the disk all energy densities remain roughly constant from $z=2$ until $z=0$. In particular, the magnetic energy density stays at $\approx 10\%$ of the thermal and kinetic energy density, well below equipartition. This can be explained by the lack of coherent, large-scale velocities in the center of the galaxy that could operate an $\alpha$-$\Omega$ dynamo and increase the magnetic energy beyond what turbulent amplification yields. Note also, that the higher densities in the center that lead to higher star formation rates also lead to significant loss of magnetic energy. In addition, the center of the galaxy generates the strongest outflows due to the high star formation rate and the AGN. In contrast to the galactic winds which are launched from the surface of the disk, the AGN outflows are launched directly from the center of the galaxy and can therefore efficiently remove highly magnetised gas and transport it into the halo and beyond. Together, outflows and the locking up of magnetic energy in stars may also be relevant for the saturation level of the magnetic field strength in the center of galaxies.

At a radius of $10\,\rm{kpc}$, the magnetic field continues to grow linearly after the initial phase of exponential amplification. The growth is stopped once the magnetic energy density reaches equipartition with the thermal energy density and the turbulent kinetic energy density at around $z=0.4$. The amplification of the magnetic field in this phase is consistent with our previous results \citep{Pakmor2014} and can be understood as the result of a galactic dynamo \citep{Shukurov2006} that is dominated by the differential rotation terms with an essentially constant radial magnetic field strength as found in our simulations (see Fig.~\ref{fig:energydirs}). The amplification of the magnetic energy is stopped when the azimuthal and vertical velocity field are not able to surpass the magnetic pressure anymore and the $\alpha$-effect is suppressed. As shown in Fig.~\ref{fig:energydirs}, the galactic dynamo leads to a magnetic field that is dominated by its azimuthal component. The resolution effects as discussed in detail in Sec.~\ref{sec:resolution} also support our picture of a galactic dynamo driving the linear amplification phase.

In addition to the amplification by the galactic dynamo, the difference between magnetic energy density and thermal and turbulent kinetic energy density is reduced by the slow overall decrease of thermal and turbulent kinetic energy density after $z=1$. Note that the time at which equipartition is reached roughly coincides with the time when the disks at this radius are finally fully developed and the total kinetic energy density does not increase any further, but completely dominates the total energy density.

The evolution at a radius of $30\,\rm{kpc}$ is similar to the evolution at $10\,\rm{kpc}$, but significantly delayed. Here, equipartition between magnetic energy density and thermal energy density is only reached around $z=0$. Note also, that some gas disks are smaller than $30\,\rm{kpc}$ and some galaxies in our sample do not have a gas disk at $z=0$ at all.

\begin{figure*}
  \centering
  \includegraphics[width=0.99\textwidth]{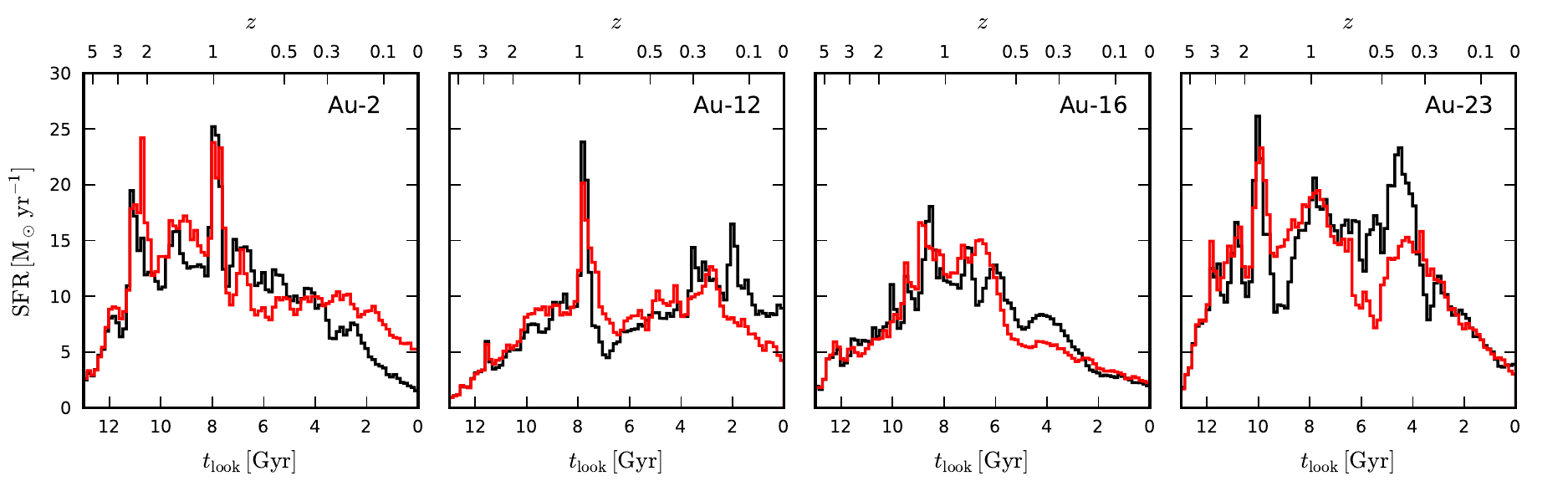}
  \caption{Comparison of the star formation history for four of the Auriga galaxies with magnetic fields (black) and without magnetic fields (red), but otherwise using an identical numerical setup.}
  \label{fig:dynamic}
\end{figure*}

\subsection{Evolution of the radial energy profiles}

The time evolution of the median energy density profiles is shown in Fig.~\ref{fig:medianrprof}. The turbulent energy density is again calculated by subtracting the Keplerian velocity before calculating the kinetic energy as described above. The radial profiles can be separated into two regimes. In the central few kpc there is little evolution between $z=1$ and $z=0$. Thermal and turbulent energy rise by more than an order of magnitude from $r = 2\,\rm{kpc}$ to the center. The magnetic energy rises less steeply and only reaches $10\%$ of the other energies as discussed above.

At the boundary between the central zone and the disk around $r=2\,\rm{kpc}$, magnetic, thermal, and turbulent energy are in equipartition already at $z=1$. With time, the part of the disk in which the magnetic energy is in equipartition increases to larger and larger radii until it extends all the way to $30 \, \rm{kpc}$ at $z=0$. This is likely a consequence of the galactic dynamo operating on the orbital timescale of the disk which is larger at larger radii. Moreover, between $z=0.5$ and $z=0$, the turbulent and thermal energy densities drop significantly in the disk, helping the magnetic energy to reach equipartition. Note also that the total kinetic energy density increases significantly at larger radii between $z=1$ and $z=0$ which reflects the formation and growth of the gas disks. 

\begin{figure*}
  \centering
  \includegraphics[width=0.99\textwidth]{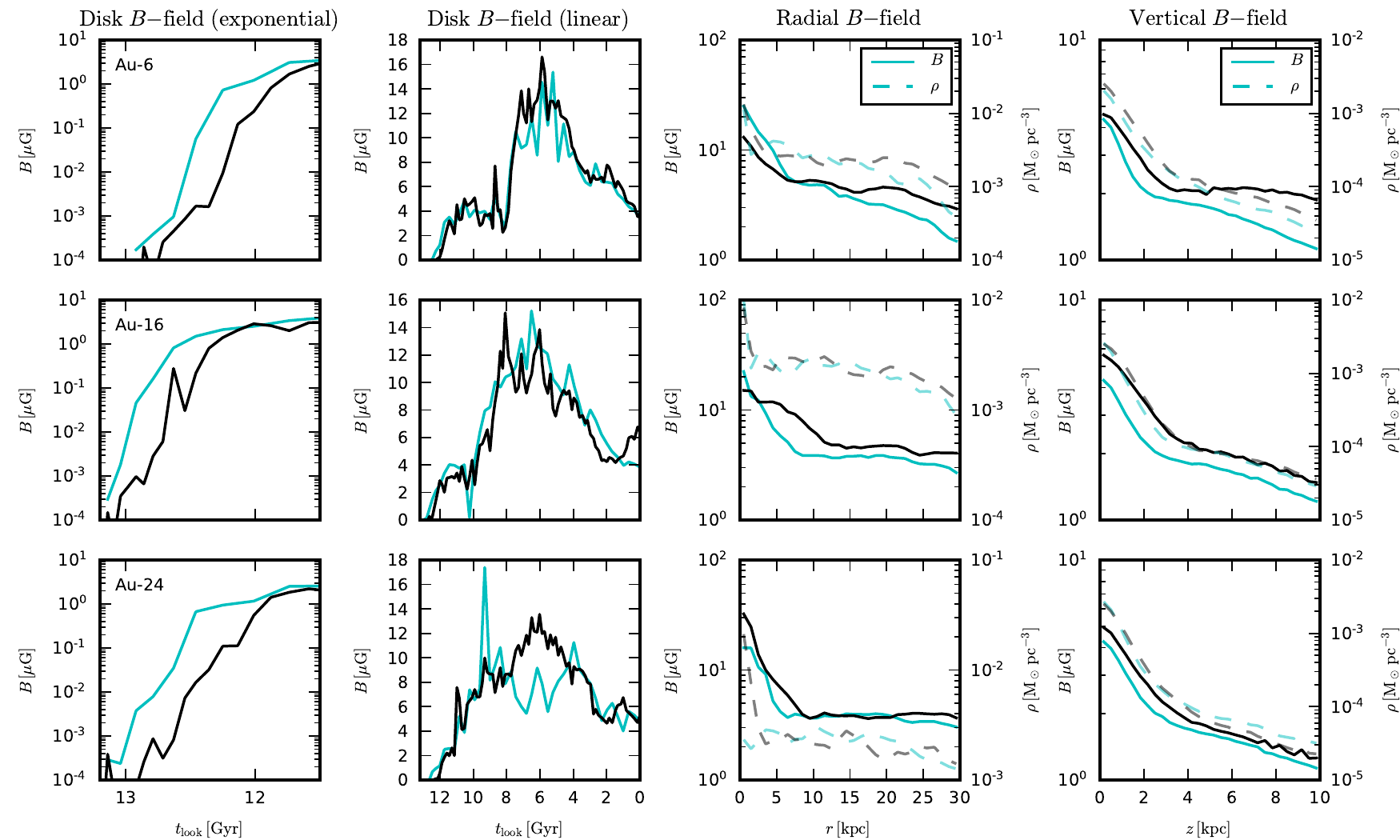}
  \caption{Detailed numerical resolution comparison for three of the Auriga halos. Black lines show the galaxies at our standard resolution, cyan lines show the same halo at `level-3' resolution, i.e.~with a factor 8 better mass resolution and half the gravitational softening. The columns show, from left to right, the early phase of the magnetic field amplification at a distance of $10\,\rm{kpc}$ from the center of the galaxy, its subsequent time evolution, the radial profile of the magnetic field strength at $z=0$, and the vertical profile of the magnetic field strength at $z=0$. The dashed lines in the third and fourth panel show the radial and vertical gas density profiles at the present epoch.}
  \label{fig:resolution}
\end{figure*}

For the median energy density profiles, the magnetic energy density can surpass the thermal energy density, but does not exceed the turbulent kinetic energy. The latter can be interpreted as a consequence of requiring a strong enough turbulent velocity component ($\alpha$-effect) to keep the galactic dynamo going. Nevertheless, for individual halos, the magnetic energy density exceeds the turbulent energy density locally, as shown for halo 6 and halo 23 in Fig.~\ref{fig:halosrprof}. However, as shown there, even in the cases when the magnetic energy exceeds the turbulent energy density, it has only decreased from the value it had when it was in equipartition with the turbulent energy density at earlier times. Therefore the magnetic energy is still only amplified until it reaches equipartition with the turbulent energy density, but the turbulent energy density then drops faster than the magnetic energy density. This leaves the magnetic energy density to dominate. Moreover, for the halos shown in Fig.~\ref{fig:halosrprof} this only happens after $z=0.2$, the redshift at which equipartition was reached throughout the disk and the magnetic energy density does not exceed the turbulent energy density by more than a factor of two. Note also that this effect likely depends on the properties of our ISM model.

\section{Dynamic impact of the magnetic field and resolution effects}
\label{sec:resolution}

Since the measurements of the magnetic fields in the Milky Way and nearby galaxies indicate that the magnetic energy is in equipartition, one may speculate whether this results in a signficant dynamical impact of magnetic fields on the evolution of galaxies. In Fig.~\ref{fig:dynamic}, we compare the star formation history of four of the Auriga galaxies with their counterpart runs without magnetic fields. At early times, the evolution is virtually identical. This is consistent with our finding that the turbulent dynamo saturates at the $10\%$ level and therefore the magnetic field is dynamically irrelevant. It only becomes dynamically relevant at much later times, when the magnetic energy density approaches the turbulent and thermal energy density in the ISM. At these later times, after $z=1$, we see deviations in the star formation history. However, these deviations are small and not systematic, i.e.~the star formation rate can increase or decrease as a result of the magnetic fields. Therefore, we conclude that in our simulations magnetic fields only have a small effect on the overall dynamic evolution of galaxies, primarily because they reach equipartition too late to affect most of the star formation.

To understand the influence of numerical resolution on our results we compare three of our halos with higher resolution counterparts in Fig.~\ref{fig:resolution}. The higher resolution runs employ $8$ times better mass resolution and $2$ times better spatial resolution. The early amplification phase that is driven by a turbulent dynamo shows the resolution dependence expected for a small-scale dynamo. At higher resolution, the dissipation scale and hence the amplification timescale become smaller, leading to a slightly faster amplification of the magnetic field. Saturation is therefore reached slightly earlier in the higher resolution runs, but it is reached at the same strength for both runs. The subsequent evolution of the magnetic field strength is then very similar and well converged with numerical resolution.

However, there are some small but systematic differences in the final magnetic field strength profiles. Except for its center, the magnetic field strength in the disk is slightly lower in the higher resolution runs. This difference is not reflected in the radial profile of the gas density and can be understood as a consequence of slightly thinner disks with smaller vertical velocities at higher resolution. The smaller vertical velocities then lead to a saturation of the dynamo at slightly lower field strength as $\alpha$-quenching sets in earlier. The vertical profile of the magnetic field strength shows the same effect, but somewhat more clearly because the vertical profile is averaged over a radius of $30\,\rm{kpc}$. Here, the slightly smaller magnetic field strength in the center also reduces the magnetic field strength in the halo, again without a corresponding difference in the gas density. Note, however, that these residual differences are only at the level of $30\%$. Nevertheless, it will be important to see how the magnetic field strength changes for even higher resolution, or when a more sophisticated model of the turbulent ISM is included.

\section{Summary and Conclusions}

\label{sec:summary}

In this study, we have analysed the time evolution and the final spatial profiles of the magnetic field strength in the Auriga disk galaxy simulations. We find that the radial and vertical profiles of the predicted magnetic field strength at $z=0$ can both be well described by two joint exponentials. Moreover, the central magnetic field strength depends on the central gas density as $B \propto \rho^{2/3}$ consistent with isotropic compression of the magnetic field of gas flowing from the disk to the center of the galaxy. The vertical profile of the magnetic field strength also strongly correlates with gas density, but exhibits a dependency of $B \propto \rho^{1/3}$. This relation is expected for adiabatic expansion of gas mostly along magnetic field lines, with only a small component of the expansion occuring perpendicular to the field. This can be naturally explained by the good alignment of the magnetic field and velocity field in the disk, and outflows happening perpendicular to the disk which reorient the magnetic field at the disk-halo interface.

Consistent with earlier exploratory work \citep{Pakmor2014}, we find that there are two distinct modes of magnetic field amplification.At high redshift, a turbulent dynamo leads to an exponential amplification of the magnetic fields in halos. We show that the turbulence is driven gravitationally by accretion of gas on the galaxy with an injection scale similar to the virial radius of the halo at $z=5$. The turbulent dynamo starts once the injection scale is sufficiently resolved with around $30$ cells. Powerspectra of kinetic and magnetic energy are consistent the expected scalings for turbulent amplification. The turbulent dynamo saturates on the smallest scales first when the magnetic energy density reaches about $20\%$ of the turbulent energy density. The magnetic dynamo sets in slightly earlier and has a slightly faster e-folding timescale in the high-resolution level 3 runs compared to the standard-resolution level 4 runs. The typical e-folding timescale in the center of the halos is $100\,\rm{Myr}$, consistent with analytic estimates for a small-scale accretion-driven dynamo in young spherical galaxies \citep{Schober2013}.

After the initial turbulent amplification phase has saturated, a second phase of magnetic field amplification starts that leads to a linear increase of the magnetic field strength with time that can be explained with a galactic dynamo with roughly constant radial magnetic field strength. This amplification only happens outside the center and is slower at larger radii, consistent with a dynamo driven by differential rotation in the disk. This second dynamo saturates when the magnetic energy reaches equipartition with turbulent and thermal energy, which happens around $z=0.5$ at a radius of $10\,\rm{kpc}$, and takes until $z=0$ for a radius of $30\,\rm{kpc}$. Reaching equipartition this late also means that despite the magnetic field being in equipartition at late times, it is largely subdominant at early times when most of the star formation takes place. Therefore, the impact of magnetic fields on the overall evolution and shape of galaxies is quite limited in our simulations.

We find that the magnetic energy density can eventually exceed the thermal and turbulent energy density at late times in the outer parts of the disk. This dominance of the magnetic energy, however, is not a result of amplifying the magnetic energy beyond the other energy components, but a consequence of the faster drop of thermal and turbulent energy density after the three energies reach equipartition. This finding may change when a more realistic treatment of non-ideal MHD effects (e.g.~reconnection) is included.

Our results are reassuringly well converged with numerical resolution, although there is a small residual trend left, with the saturation level of the magnetic field strength in the disk becoming smaller with increasing resolution, reflecting the fact that the disks get a bit colder with improved resolution. We note that none of our physics model parameters are changed when the numerical resolution is varied.

Finally, we find that the magnetic field reaches equipartition quite late, which in turn implies that it is dynamically irrelevant for most of the time. Therefore, including magnetic fields has little direct influence on the evolution of our Milky Way-like galaxies, as for example seen in their star formation history. However, this picture may well change substantially when anisotropic transport processes are included that directly depend on the orientation of the magnetic field \citep{Pakmor2016b}. Here the magnetic field can play a critical role for the evolution of galaxies even if its pressure is negligible. This alone provides ample motivation to further study the origin of magnetic fields in galaxies, in particular when such transport processes of thermal energy, momentum, or cosmic rays are included as well.
    
\section*{Acknowledgements}
    
This work has been supported by the European Research Council under ERC-StG grant EXAGAL-308037, ERC-CoG grant CRAGSMAN-646955, and by the Klaus Tschira Foundation. RP, RG and VS acknowledge support by the DFG Research Centre SFB-881 `The Milky Way System` through project A1. VS and TG acknowledge support through subproject EXAMAG of the Priority Programme 1648 `Software for Exascale Computing' of the German Science Foundation. DJRC acknowledges the support of STFC studentship ST/K501979/1. This work used the Data Centric system at Durham University, operated by the Institute for Computational Cosmology on behalf of the STFC DiRAC HPC Facility ``www.dirac.ac.uk``. This equipment was funded by BIS National E-infrastructure capital grant ST/K00042X/1, STFC capital grant ST/H008519/1, and STFC DiRAC Operations grant ST/K003267/1 and Durham University. DiRAC is part of the UK National E-Infrastructure.
    
\bibliographystyle{mnras}

\label{lastpage}

\end{document}